\documentclass{vldb}
\usepackage{xspace}
\usepackage{color}
\usepackage[normalem]{ulem}

\usepackage{balance}

\usepackage{multibib}

\usepackage{longtable}

\newcites{app}{Appendix}

\newcommand{\dimm}{\textsc{DimmWitted}\xspace}
\newcommand{\hogwild}{Hogwild!\xspace}
\newcommand{\graphchi}{GraphChi\xspace}
\newcommand{\graphlab}{GraphLab\xspace}
\newcommand{\spark}{Spark\xspace}
\newcommand{\mllib}{MLlib\xspace}

\newcommand{\percpu}{\textsf{PerCore}\xspace}
\newcommand{\permachine}{\textsf{PerMachine}\xspace}
\newcommand{\pernode}{\textsf{PerNode}\xspace}

\newcommand{\sharding}{\textsf{Sharding}\xspace}
\newcommand{\fullreplication}{\textsf{FullReplication}\xspace}
\newcommand{\OS}{\textsf{OS}\xspace}
\newcommand{\NUMA}{\textsf{NUMA}\xspace}
\newcommand{\Dense}{\textsf{Dense}\xspace}
\newcommand{\Sparse}{\textsf{Sparse}\xspace}
\newcommand{\Rowmajor}{\textsf{Row-major}\xspace}
\newcommand{\Columnmajor}{\textsf{Column-major}\xspace}

\newcommand{\R}{\mathbb{R}}

\newtheorem{example}{Example}[section]

\newcommand{\torevise}[1]{{#1}}

\newcommand{\yell}[1]{{#1}}

\newcommand{\eat}[1]{}

\def\compactify{\itemsep=-5pt \topsep=-5pt \partopsep=-2pt \parsep=-5pt}
\let\latexusecounter=\usecounter

\newenvironment{CompactItemize}
 {\def\usecounter{\compactify\latexusecounter}
  \compactify\begin{itemize}}
 {\end{itemize}\let\usecounter=\latexusecounter}

\usepackage{algpseudocode}
\usepackage{algorithm}
\usepackage{algorithmicx}
\usepackage{multirow}
\usepackage{url}
\usepackage{makecell}
\usepackage{enumitem,cite}

\begin{document}
\title{DimmWitted: A Study of Main-Memory Statistical Analytics}

\author{
\alignauthor
Ce Zhang$^{\dagger\ddagger}$~~~~~~~~~~Christopher R\'{e}$^\dagger$\\
\affaddr{$^\dagger$Stanford University}\\
\affaddr{$^\ddagger$University of Wisconsin-Madison}
\email{\{czhang, chrismre\}@cs.stanford.edu}
}
\maketitle

\begin{abstract}

We perform the first study of the tradeoff space of access methods and
replication to support statistical analytics using first-order methods
executed in the main memory of a Non-Uniform Memory Access (NUMA)
machine. Statistical analytics systems differ from conventional
SQL-analytics in the amount and types of memory incoherence 
\yell{that} they can
tolerate.  Our goal is to understand tradeoffs in accessing the data
in row- or column-order and at what granularity one should share the
model and data for a statistical task. We study this new tradeoff
space and discover \yell{that} there are tradeoffs between 
hardware and
statistical efficiency. We argue that our tradeoff study may provide
valuable information for designers of analytics engines: for each
system we consider, our prototype engine can run at least one popular
task at least 100$\times$ faster. We conduct our study across five
architectures using popular models\yell{,} including SVMs, logistic
regression, Gibbs sampling, and neural networks.
\end{abstract}

\begin{figure*}[t!]
\centering
\includegraphics[width=0.95\textwidth]{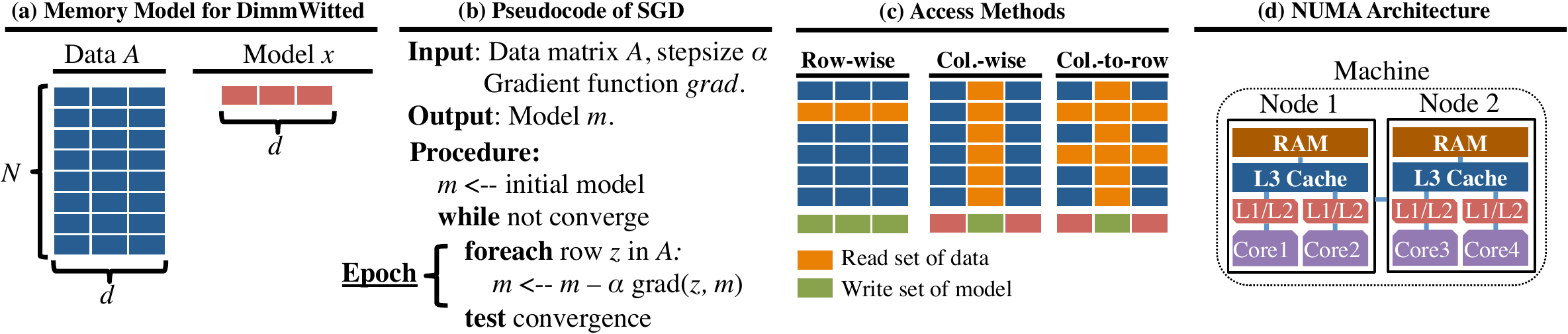}
\caption{Illustration of (a) \dimm's Memory Model, (b) Pseudocode for SGD,
  (c) Different Statistical Methods in \dimm and Their Access Patterns,
  and (d) NUMA Architecture.}
 \label{fig:basic_block}
 \end{figure*}

\section{Introduction}

Statistical data analytics is one of the hottest topics in
data-management research and practice. Today, even small organizations
have access to machines with large main memories (via Amazon's EC2) or
for purchase at \$5/GB. As a result, there has been a flurry of
activity to support main-memory analytics in both industry (Google
Brain, Impala, and Pivotal) and research (GraphLab, and MLlib). Each
of these systems picks one design point in a larger tradeoff space.
The goal of this paper is to define and explore this space. We find
that today's research and industrial systems under-utilize
\yell{modern commodity} hardware for analytics---sometimes by two orders of
magnitude. We hope that our study identifies some useful design points
for the next generation of such main-memory analytics systems.

Throughout, we use the term {\em statistical analytics} to refer to
those tasks that can be solved by {\em first-order methods}--a class
of iterative algorithms that use gradient information; these methods
are the core algorithm in systems \yell{such as} MLlib, GraphLab, and
Google Brain.
Our study examines analytics on commodity multi-socket, multi-\torevise{core},
non-uniform memory access (NUMA) machines, which are the de facto
standard machine configuration and \yell{thus} a natural target 
for an in-depth study. 
Moreover, our experience with several enterprise companies
suggests that, after appropriate preprocessing, a large class of
enterprise analytics problems fit into the main memory of a single,
modern machine. While this architecture has been recently studied for
traditional SQL-analytics systems~\cite{Chasseur:2013:VLDB}, it has
not been studied for {\it statistical} analytics
systems. 

Statistical analytics systems are different from traditional
SQL-analytics systems. In comparison to traditional SQL-analytics, the
underlying methods are intrinsically robust to error. On the other
hand, traditional statistical theory does not consider which
operations can be efficiently executed. This leads to a fundamental
tradeoff between {\em statistical efficiency} (how many steps are
needed until convergence to a given tolerance) and {\em hardware
  efficiency} (how efficiently those steps can be carried out).

To describe such tradeoffs more precisely, we describe the setup of
the analytics tasks that we consider in this paper. The input data is
a matrix in $\R^{N \times d}$ and the goal is to find a vector $x \in
\R^{d}$ that minimizes some (convex) loss function, say the logistic
loss or \yell{the} hinge loss (SVM). Typically, one makes several complete passes
over the data while updating the model; we call each such pass an {\em
  epoch}. There may be some communication at the end of the epoch,
e.g., in bulk-synchronous parallel systems \yell{such as} Spark. We
identify three tradeoffs that have not been explored in the
literature: (1) {\it access methods for the data}, (2) {\it model
  replication}, and (3) {\it data replication}. Current systems have
picked one point in this space; we explain each space and discover
points that have not been previously considered. Using these new
points, we can perform 100$\times$ faster than previously explored
points in the tradeoff space for several popular tasks.

\paragraph*{Access Methods} 
Analytics systems access (and store) data in either row-major or
column-major order. For example, systems that use {\it stochastic
  gradient \yell{descent} methods} (SGD) access the data row-wise; examples include
MADlib~\cite{Hellerstein:2012:VLDB} in Impala and Pivotal, Google
Brain~\cite{DBLP:conf/icml/LeRMDCCDN12}, and MLlib in
Spark~\cite{Sparks:2013:ICDM}; and {\it stochastic coordinate descent
\yell{methods}} (SCD)
access the data column-wise; examples include
GraphLab~\cite{DBLP:journals/pvldb/LowGKBGH12},
Shogun~\cite{Sonnenburg:2010:JMLR}, and
Thetis~\cite{DBLP:conf/nips/SridharWRLBZ13}. These methods have
essentially identical statistical efficiency, but their wall-clock
performance can be radically different due to hardware
efficiency. However, this tradeoff has not been systematically
studied. To study this tradeoff, we introduce a storage abstraction
that captures the access patterns of popular statistical analytics
tasks and a prototype called \dimm. In particular, we identify three
access methods that are used in popular analytics tasks\yell{,} including
standard supervised machine learning models \yell{such as} SVMs, logistic
regression, \yell{and} least squares; and more advanced methods \yell{such as} neural
networks and Gibbs sampling on factor graphs. For different access
methods for the same problem, we find that the time to converge to a
given loss can differ by up to 100$\times$.

We also find that no access method dominates all others, \yell{so} an
engine designer may want to include both access methods. To show that
it may be possible to support both methods in a single engine, we
develop a simple cost model to choose among these access methods. We
describe a simple cost model that selects a nearly optimal point in
our data sets, models, and different machine configurations.

\paragraph*{Data and Model Replication}
We study two sets of tradeoffs: the level of granularity, and the
mechanism by which mutable state and immutable data are shared
in analytics tasks. We describe the tradeoffs we explore in both (1)
mutable state sharing, which we informally call {\em model
  replication}, and (2) {\em data replication}.

\paragraph*{(1) Model Replication}
During execution, there is some state that the task mutates (typically
an update to the model). We call this state, which may be shared among
one or more processors, a {\em model replica}. We consider three
different granularities at which to share model replicas:

\begin{CompactItemize}
\item The \percpu approach treats a NUMA machine as a
  distributed system in which every \torevise{core} is treated as an individual
  machine, e.g., in bulk-synchronous models \yell{such as} MLlib on Spark or
  event-driven systems \yell{such as} GraphLab. These approaches are the
  classical shared-nothing and event-driven architectures,
  respectively. In \percpu, the part of the model that is updated by
  each \torevise{core} is only visible to that \torevise{core} until the end of an
  epoch. This method is efficient and scalable from a hardware
  perspective, but it is less statistically efficient\yell{,} 
  as there is only coarse-grained communication between \torevise{cores}.

\item The \permachine approach acts as if each processor has uniform
  access to memory. This approach is taken in Hogwild! and Google
  Downpour~\cite{DBLP:conf/nips/DeanCMCDLMRSTYN12}. In this method,
  the hardware takes care of \yell{the} coherence of the shared state. The
  \permachine method is statistically efficient due to high
  communication rates, but it may cause contention in the hardware,
  which may lead to suboptimal running times.

\item A natural hybrid is \pernode; this method uses the fact that
  \percpu communication through the last-level cache (LLC) is
  dramatically faster than communication through remote main
  memory. This method is novel; for some models, \pernode can be an
  order of magnitude faster.
\end{CompactItemize}

\noindent
Because model replicas are mutable, a key question is \yell{{\it how
    often should we synchronize model replicas?}} We find that it is
beneficial to synchronize the models as much as possible---so long as
we do not impede throughput to data in main memory. A natural idea,
then, is to use \permachine sharing, in which the hardware is
responsible for synchronizing the replicas. However, this decision can
be suboptimal\yell{,} as the cache-coherence protocol may stall a
processor to preserve coherence\yell{,} but this information may not
be worth the cost of a stall from a statistical efficiency
perspective. We find that the \pernode method, coupled with a simple
technique to batch writes across sockets, can dramatically reduce
communication and \torevise{processor} stalls. The \pernode method can
result in an over $10\times$ runtime improvement. This technique
depends on the fact that we do not need to maintain the model
consistently: we are effectively delaying some updates to reduce the
total number of updates across sockets (which lead to processor
stalls).

\paragraph*{(2) Data Replication} The data for analytics is immutable, so there
are no synchronization issues for data replication. The classical
approach is to partition the data to take advantage of higher
aggregate memory bandwidth. However, each partition may contain skewed
data, which may slow convergence. Thus, an alternate approach
is to replicate the data \yell{fully} (say, per NUMA node). In this approach,
each node accesses that node's data in a different order, which means
that the replicas provide non-redundant statistical information; in
turn, this reduces the variance of the estimates based on the data in
each replicate. We find that \yell{for} some tasks, fully replicating the data
four ways can converge to the same loss almost 4$\times$ faster than
the sharding strategy.

\paragraph*{Summary of Contributions} 
We are the first to study the three tradeoffs \yell{listed} 
above for main-memory
statistical analytics systems. These tradeoffs are not intended to be
an exhaustive set of optimizations, but they demonstrate our main
conceptual point: {\it treating \yell{NUMA machines} as distributed systems
  or SMP is suboptimal for statistical analytics}. We design a storage
manager, \dimm, that shows it is possible to exploit these ideas on
real data sets. Finally, we evaluate our techniques on multiple real
datasets, models, and architectures.


\section{Background} \label{sec:background}

\begin{sloppypar}
In this section,
we describe the memory model for \dimm, which provides a unified
memory model to implement popular analytics methods. Then, we recall
some basic properties of modern NUMA architectures.
\end{sloppypar}

\paragraph*{Data for Analytics}
The data for an analytics task is a pair $(A,x)$, which we call the
data and the model\yell{,} respectively. For concreteness, we consider a
matrix $A \in \R^{N \times d}$. In machine learning parlance, each row
is called an {\em example}. Thus, $N$ is often the number of examples
and $d$ is often called the dimension of the model. There is also a
model, typically a vector $x \in \R^{d}$. The distinction is that
the data $A$ is read-only\yell{,} while the model vector, $x$, will be updated
during execution. From the perspective of this paper, the important
distinction we make is that data is an immutable matrix\yell{,} while the model
(or portions of it) are mutable data.

\paragraph*{First-Order Methods for Analytic Algorithms} 
\dimm considers a class of popular algorithms called {\em first-order
  methods}. Such algorithms make several passes over the data; we
refer to each such pass as an {\em epoch}. A popular example algorithm
is \yell{stochastic gradient descent} (SGD), which is widely used \yell{by}
web-companies, e.g., Google Brain~\cite{DBLP:conf/icml/LeRMDCCDN12}
and VowPal Wabbit~\cite{DBLP:journals/corr/abs-1110-4198}, and in
enterprise systems \yell{such as} Pivotal, Oracle, and Impala. Pseudocode for
this method is shown in Figure~\ref{fig:basic_block}(b). During each
epoch, SGD reads a single example $z$; it uses the current value of
the model and $z$ to estimate the derivative; \yell{and}
it then updates the
model vector with this estimate. It reads each example in this
loop. After each epoch, these methods test convergence (usually by
computing or estimating the norm of the gradient); this computation
requires a scan over the complete dataset.

\subsection{Memory Models for Analytics}

We design \dimm's memory model to capture the trend in recent 
\yell{high-performance} sampling and statistical methods. There are two aspects to
this memory model: the {\em coherence level} and the {\em storage
  layout}.

\paragraph*{Coherence Level} 
Classically, memory systems are coherent: reads and writes are
executed atomically. For analytics systems, we say \yell{that}
a memory model is
{\em coherent} if reads and writes of the entire model vector are
atomic. That is, access to the model is enforced by a critical
section. However, many modern analytics algorithms are designed for an
{\em incoherent} memory model. The Hogwild! method showed that one can
run such a method in parallel without locking but still provably
converge. The Hogwild!  memory model relies on the fact that writes of
individual components are atomic, but it does not require that
the entire vector be updated atomically. However, atomicity at the
level of the cacheline is provided by essentially all modern
processors. Empirically, these results allow one to forgo costly
locking (and coherence) protocols. Similar algorithms have been proposed
for other popular methods\yell{,} including Gibbs
sampling~\cite{Smola:2010:VLDB,Johnson:2013:NIPS}, 
\yell{stochastic coordinate descent}
(SCD)~\cite{Richtarik:2012:ArXiv,Sonnenburg:2010:JMLR}, and linear
systems solvers~\cite{DBLP:conf/nips/SridharWRLBZ13}. This technique
\yell{was} applied by Dean et
al.~\cite{DBLP:conf/nips/DeanCMCDLMRSTYN12} to solve convex
optimization problems with billions of elements in a model. This memory model
is distinct from the classical, {\em fully coherent} database
execution.

The \dimm prototype allows us to specify that a region of memory is
coherent or not. This region of memory may be shared by one or more
processors. If the memory is only shared per thread, then we can
simulate a shared-nothing execution. If the memory is shared per
machine, we can simulate Hogwild!.

\begin{figure}[t!]
\centering
\includegraphics[width=0.5\textwidth]{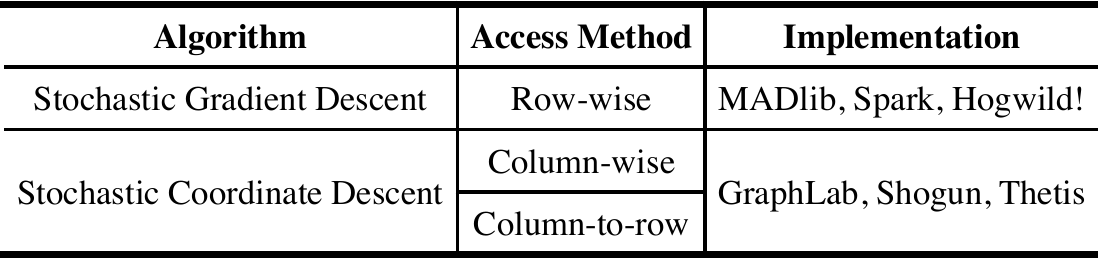}
\caption{Algorithms and Their Access Methods.}
 \label{fig:access_methods}
 \end{figure}


\paragraph*{Access Methods} 

We identify three distinct access paths used by modern
analytics systems, which we call row-wise, column-wise, and
column-to-row. They are graphically illustrated in
Figure~\ref{fig:basic_block}(c). Our prototype supports all three access
methods. All of our methods perform several epochs, that is, passes over the
data. However, the algorithm may iterate over the data row-wise or
column-wise.

\begin{itemize}
\begin{sloppypar}
\item In {\em row-wise access}, the system scans each row of the table
  and applies a function that takes that row, applies a function to
  it, and then updates the model. This method may write to all
  components of the model. Popular methods that use this access method
  include stochastic gradient descent, gradient descent, and 
  \yell{higher-order} methods (\yell{such as} l-BFGS). 
\end{sloppypar}

\item In {\em column-wise access}, the system scans each column $j$ of
  the table. This method reads just the $j$ component of the
  model. The write set of the method is typically a single component
  of the model. This method is used by stochastic coordinate descent.

\item In {\em column-to-row access}, the system iterates conceptually
  over the columns. This method is typically applied to sparse
  matrices. When iterating on column $j$\yell{,} it will read all rows in
  which column $j$ is non-zero. This method also updates a single
  component of the model. This method is used by non-linear support
  vector machines in \graphlab and is the de facto approach for Gibbs
  sampling.

\end{itemize}
\dimm is free to iterate over rows or columns in essentially any order
(although typically some randomness in the ordering is
desired). Figure~\ref{fig:access_methods} classifies popular
implementations by their access method.

\subsection{Architecture of NUMA Machines}

\begin{figure}[t!]
\centering
\includegraphics[width=0.48\textwidth]{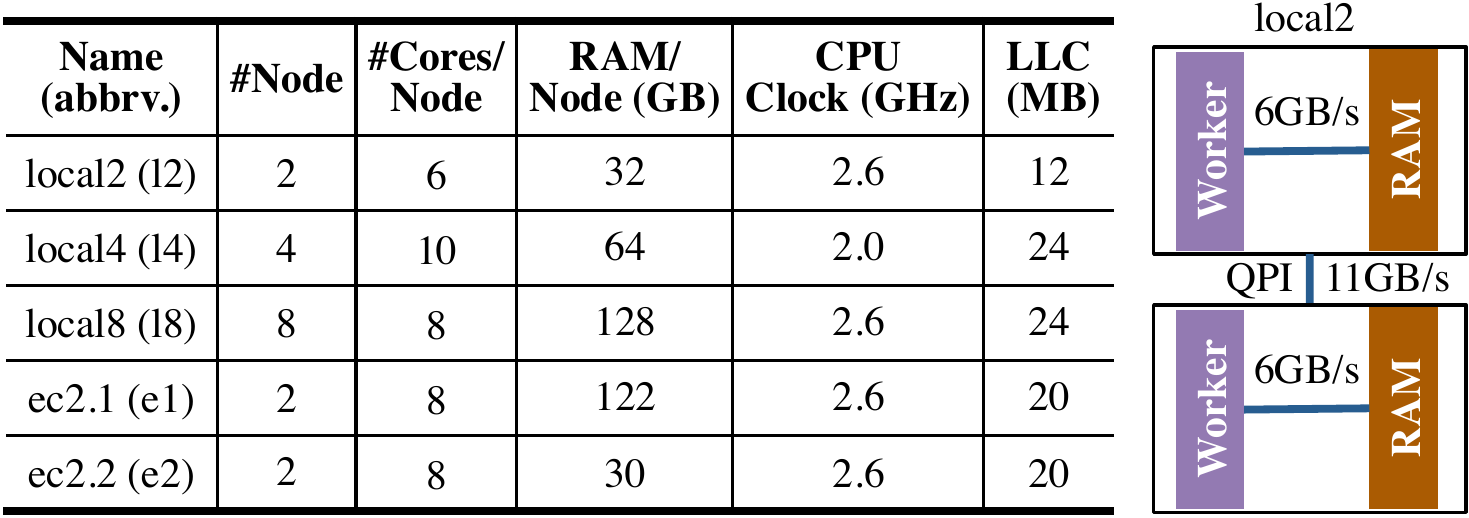}
\caption{Summary of Machines and Memory Bandwidth on local2 Tested with STREAM~\protect\cite{Bergstrom:2011:ArXiv}.}
 \label{fig:architecture}
 \end{figure}

We briefly describe the architecture of a modern NUMA machine.
As illustrated in Figure~\ref{fig:basic_block}(d), a NUMA machine
contains multiple NUMA nodes. Each node has multiple cores and
processor caches, including the L3 cache. Each node is directly
connected to a region of DRAM. NUMA nodes are connected to each other
by buses on the main board; in our case, this connection is the Intel
Quick Path Interconnects (QPIs), which has a bandwidth as high as
25.6GB/s.\footnote{\url{www.intel.com/content/www/us/en/io/quickpath-technology/quick-path-interconnect-introduction-paper.html}}
To access DRAM regions of other NUMA nodes, data is transferred across
NUMA nodes using the QPI. These NUMA architectures are cache coherent,
and the coherency actions use the QPI. Figure~\ref{fig:architecture}
describes the configuration of each machine that we use in this
paper. Machine\yell{s} controlled by us have names with the prefix ``local'';
the other machines are Amazon EC2 configurations.

\begin{figure}[t]
\centering
\includegraphics[width=0.35\textwidth]{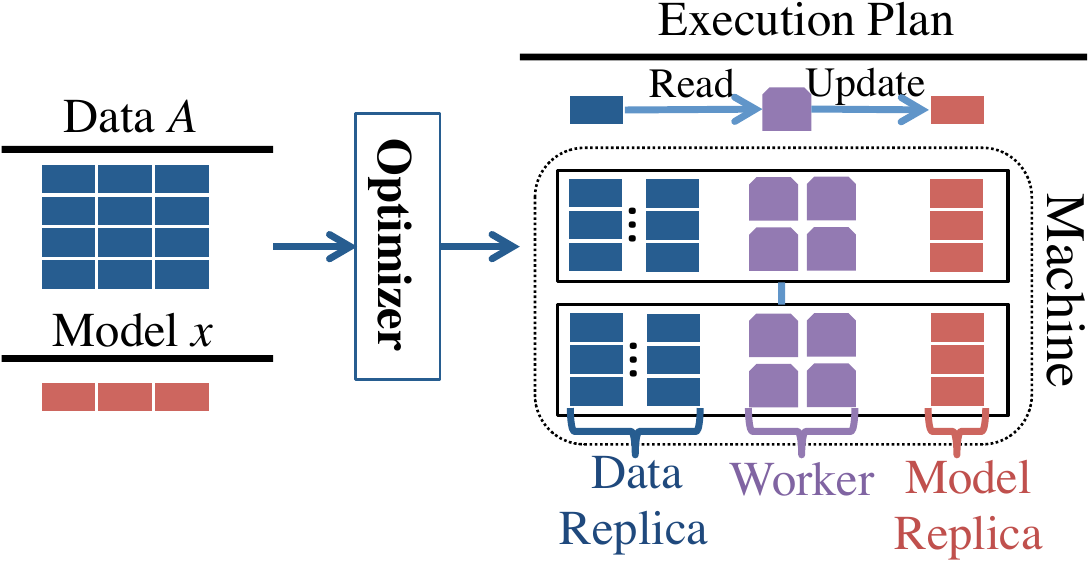}
\caption{Illustration of \dimm's Engine.}
\label{fig:system_overview}
\end{figure}


\section{The DimmWitted Engine } \label{sec:optimizer}

\begin{figure}[t]
\centering
\includegraphics[width=0.3\textwidth]{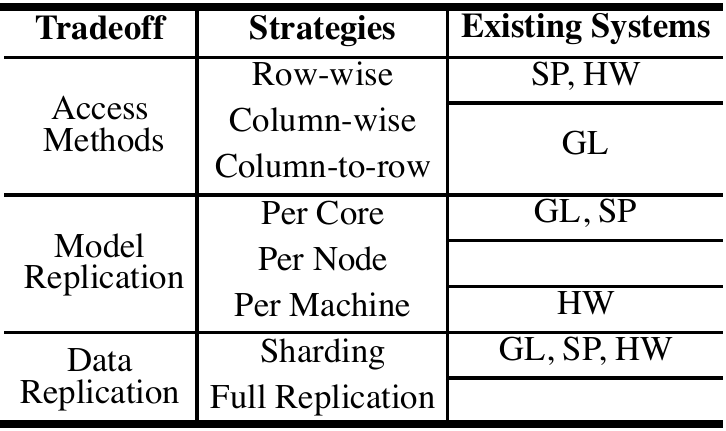}
\caption{A Summary of \dimm's Tradeoffs and
Existing Systems (GraphLab (GL), Hogwild! (HW), Spark (SP)).}
\label{fig:tradeoff}
\end{figure}

We describe the tradeoff space that \dimm's optimizer considers,
namely (1) \yell{access method selection}, (2) \yell{model replication}, 
and (3) \yell{data replication}. 
To help understand the statistical-versus-hardware
tradeoff space, we present some experimental results in a {\em
  Tradeoffs} paragraph within each subsection.  
\torevise{We describe implementation details for \dimm in the full
version of this paper.}

\subsection{System Overview} \label{sec:overview}
We describe analytics tasks in \dimm and the execution model of
\dimm given an analytics task.

\paragraph*{System Input} 
For each analytics task that we study, we assume \yell{that}
the user provides
data $A \in \R^{N \times d}$ and an initial model that is a vector of
length $d$. In addition, for each access method listed above, there is
a function of an appropriate type that solves the same underlying
model.  For example, we provide both a row- and column-wise way of
solving a support vector machine.
Each method takes two
arguments; the first is a pointer to a model.
\begin{CompactItemize}
\item $f_{row}$ captures the the row-wise access method, and its second
  argument is the index of a single row.

\item $f_{col}$ captures the column-wise access method, and its second
  argument is the index of a single column. 
  
\item $f_{ctr}$ captures the column-to-row access method, and its
  second argument is a pair of one column index and a set of row
  indexes. These rows correspond to the non-zero entries in a data
  matrix for a single column.\footnote{Define $S(j) = \{ i : a_{ij}
    \neq 0\}$. For a column $j$, the input to $f_{ctr}$ is a pair
    $(j,S(j))$.}
\end{CompactItemize}
\noindent Each of the functions \yell{modifies} the model to which they receive
a pointer in place. However, in our study\yell{,} $f_{row}$ can modify the
whole model, while $f_{col}$ and $f_{ctr}$ only modify a single
variable of the model. We call the above tuple of functions a {\em
  model specification}. Note that a model specification contains
either $f_{col}$ or $f_{ctr}$ but typically not both.

\paragraph*{Execution} 
Given a model specification, our goal is to generate an execution
plan. An execution plan, schematically illustrated in
Figure~\ref{fig:system_overview}, specifies three things for each 
CPU core in the machine: (1) a subset of the data matrix to operate on,
(2) a replica of the model to update, and (3) the access method used to
update the model. We call the set of replicas of data and models {\em locality
  groups}\yell{,} as the replicas are described physically\yell{;} i.e., they
correspond to regions of memory that are local to particular NUMA
nodes, and one or more workers may be mapped to each locality
group. The data assigned to distinct locality groups may overlap. We
use \dimm's engine to explore three tradeoffs:
\begin{description}
\itemsep=-3pt \topsep=0pt \partopsep=-2pt \parsep=-5pt
\item \yell{{\bf (1) Access methods,}} in which we can select between either
  the row or column method to access the data.

\item \yell{{\bf (2) Model replication,}} in which we choose how to create and
  assign replicas of the model to each worker. When a worker needs to
  read or write the model, it will read or write the model replica
  that it is assigned.

\item \yell{{\bf (3) Data replication,}} in which we choose a subset of data
  tuples for each worker. The replicas may be overlapping, disjoint,
  or some combination. 

\end{description}

\noindent
Figure~\ref{fig:tradeoff} summarizes the tradeoff space. In each section, we
illustrate the tradeoff along two axes, namely (1) the {\it
  statistical efficiency}, i.e., the number of epochs it takes to
converge\yell{,} and (2) {\em hardware efficiency}, the time that each method
takes to finish a single epoch.

\begin{figure}[t]
\centering
\includegraphics[width=0.48\textwidth]{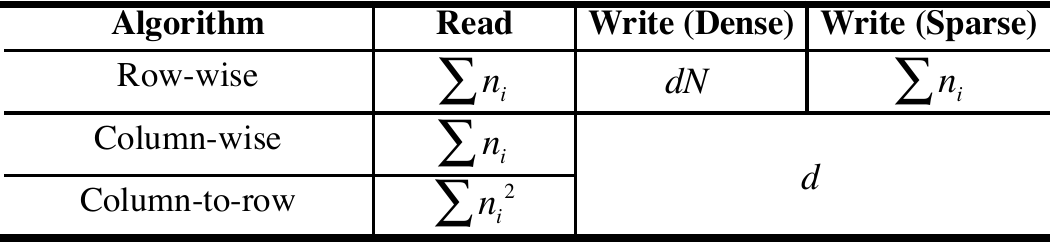}
\caption{Per Epoch Execution Cost of Row- and Column-wise Access.
\small The Write column is for a single model replica.
Given a dataset $A\in \mathbb{R}^{N\times d}$, let $n_i$ be the number of non-zero elements $a_i$.}
\label{fig:costmodel}
\end{figure}

\begin{figure}[t]
\centering
\includegraphics[width=0.4\textwidth]{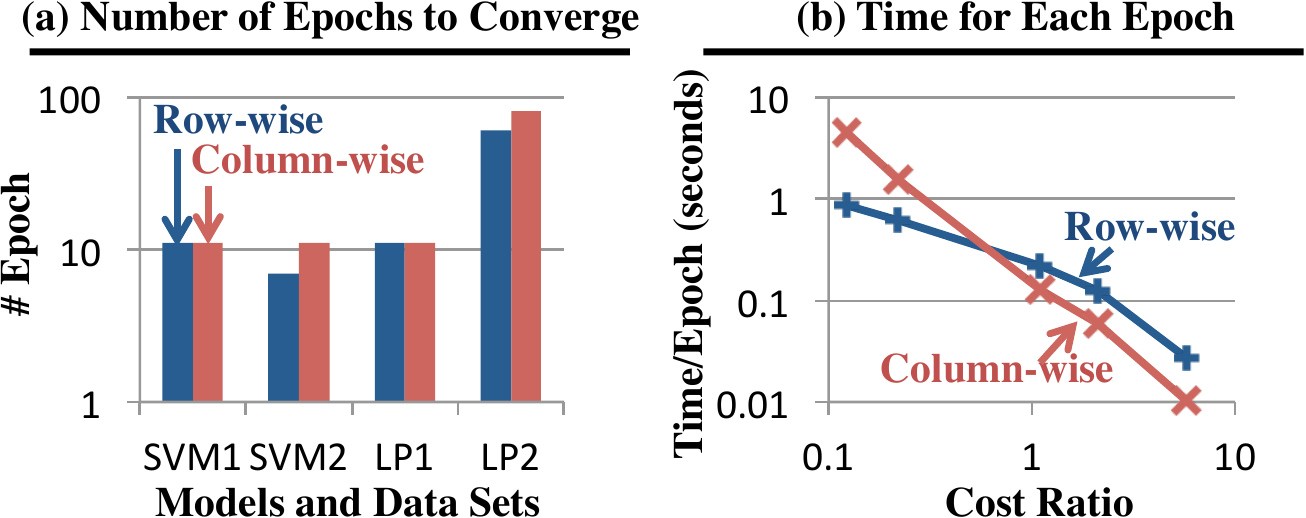}
\caption{Illustration of the Method Selection Tradeoff.  \small (a)
  These four datasets are RCV1, Reuters, Amazon, and Google,
  respectively.  (b) The \yell{``cost ratio''} is defined as the ratio of
  costs estimated for row-wise and column-wise methods:
  $(1+\alpha)\sum_i {n_i}/(\sum_i {n_i^2} + \alpha d)$, where $n_i$ is
  the number of non-zero elements of $i^{th}$ row of $A$ and $\alpha$
  is the cost ratio between writing and reads.  We set $\alpha=10$ to
  plot this graph.}
\label{fig:synthetic_sgd_scd}
\end{figure}

\subsection{Access Method Selection} \label{sec:method_selection}

In this section,
we examine each access method: row-wise, column-wise, and
column-to-row. We find that the execution time of an access method
depends more on hardware efficiency than \yell{on} statistical efficiency.

\paragraph*{Tradeoffs} 
We consider the two tradeoffs that we use for a simple cost model
(Figure~\ref{fig:costmodel}). Let $n_i$ be the number of non-zeros in
row $i$\yell{;} \torevise{when we store the data as sparse vectors/matrices
in CSR format},
the number of reads in a row-wise access method is
$\sum_{i=1}^{N} n_i$. Since each example is likely to be written back
in a dense write, we perform $dN$ writes per epoch. Our cost model
combines these two costs linearly with a factor $\alpha$ that accounts
for writes being more expensive\yell{,} on average, \yell{because of} contention. The
factor $\alpha$ is estimated at installation time by measuring on a
small set of datasets. The parameter $\alpha$ is in $4$ to $12$ and
grows with the number of sockets\yell{;} e.g., for local2, $\alpha\approx 4$,
and \yell{for} local8, $\alpha\approx 12$. Thus, $\alpha$ may increase in the future.

\begin{sloppypar}
{\bf Statistical Efficiency.} We observe that each access method
has comparable {\em statistical efficiency}. To illustrate this, we
run all methods on all of our datasets and report the number of
epochs that one method converges to a given error to the optimal loss,
and Figure~\ref{fig:synthetic_sgd_scd}(a) shows the result on four
datasets with 10\% error. We see that the gap \yell{in the} number of epochs
\yell{across} different methods \yell{is} small (always within $50\%$ of each
other).  
\end{sloppypar}

\begin{sloppypar}
{\bf Hardware Efficiency.}
Different access methods can change the time per epoch by up to a
factor of $10\times$, and there is a cross-over point. To see this, we
run both methods on a series of synthetic datasets where we control
the number of non-zero elements per row by subsampling each row on the
Music dataset (see Section~\ref{sec:experiment} for more details).
For each subsampled dataset, we plot the cost ratio on the $x$-axis,
and we plot their actual running time per epoch in
Figure~\ref{fig:synthetic_sgd_scd}(b).  We see a cross-over point on
the time used per epoch: when the cost ratio is small, row-wise
outperforms column-wise by $6\times$\yell{,} as the column-wise method reads more
data; on the other hand, when the ratio is large, the column-wise
method outperforms the row-wise method by $3\times$\yell{,} as the column-wise
method has lower write contention. We observe similar cross-over
points on our other datasets.
\end{sloppypar}

\paragraph*{Cost-based Optimizer} 
\dimm estimates the execution time of different access methods using
the number of bytes that each method reads and writes in one epoch,
as shown in Figure~\ref{fig:costmodel}. For writes, it is slightly more
complex: for models \yell{such as} SVM, each gradient step in row-wise access
only updates the coordinates where the input vector contains non-zero
elements. We call this scenario a {\em sparse} update; otherwise\yell{,}
it is a {\em dense} update. 

\dimm needs to estimate the ratio of the cost of reads to writes. To
do this, it runs a simple benchmark dataset. We find that\yell{, for all 
the} eight datasets, five statistical models, and five machines that we
use in the experiments, the cost model is robust to this parameter:
as long as writes are $4\times$ to $100\times$ more expensive than
reading, the cost model makes the correct decision between row-wise
and column-wise access.

\subsection{Model Replication} \label{sec:model_allocation}

\begin{sloppypar}
In \dimm, we consider three model replication strategies. The first
two strategies, namely \percpu and \permachine, are
similar to traditional shared-nothing and shared-memory architecture,
respectively.  We also consider a hybrid strategy, \pernode\yell{,}
designed for NUMA machines.
\end{sloppypar}


\subsubsection{Granularity of Model Replication}

The difference between the three model replication strategies is the
granularity of replicating a model.  We first describe \percpu and
\permachine and their relationship with other existing systems
(Figure~\ref{fig:tradeoff}).  We then describe \pernode, a simple,
novel hybrid strategy that we designed to leverage the
structure of NUMA machines.

\paragraph*{\percpu} In the 
\percpu strategy, each core maintains a mutable state, and these states
are combined to form a new version of the model (typically at the end
of each epoch).  This is essentially a shared-nothing architecture; it
is implemented in Impala, Pivotal, and Hadoop-based frameworks.
\percpu is popularly implemented by state-of-the-art statistical
analytics frameworks \yell{such as} Bismarck, \spark, and \graphlab. There
are subtle variations to this approach: in Bismarck's implementation,
each worker processes a partition of the data, and its model is
averaged at the end of each epoch; \spark implements a minibatch-based
approach in which parallel workers calculate the gradient based on
examples, and then gradients are aggregated by a single thread to
update the final model; \graphlab implements an event-based approach
where each different task is dynamically scheduled to satisfy the
given consistency requirement.
In \dimm, we implement \percpu in a way that is similar to
Bismarck, where each worker has its own model replica, and each worker
is responsible for updating its replica.\footnote{We
  implemented MLlib's minibatch in \dimm. We find that the
  \hogwild-like implementation always dominates the minibatch
  implementation. \dimm's column-wise implementation for \permachine
  is similar to \graphlab, with the only difference that \dimm does
  not schedule the task in an event-driven way. } As we will show in
the experiment section, \dimm's implementation is 3-100$\times$ faster
than either \graphlab and \spark. Both systems have additional sources
of overhead that \dimm does not, e.g., for fault tolerance in \spark
and a distributed environment in both. We are not making an argument
about the relative merits of these features in applications, only that
they would obscure the tradeoffs that we study in this paper.

\paragraph*{\permachine} In the \permachine strategy, there is 
a single model replica that all workers update during execution.
\permachine is implemented in \hogwild and Google's Downpour. \hogwild
implements a lock-free protocol, which forces the hardware to deal
with coherence. Although different writers may overwrite each other
and readers may have dirty reads, Niu et
al.~\cite{DBLP:conf/nips/RechtRWN11} prove that \hogwild converges.

\paragraph*{\pernode} The \pernode strategy
is a hybrid of \percpu and \permachine. In \pernode,
each NUMA node has a single model replica that is shared among all
\torevise{cores} on that node.

\paragraph*{Model Synchronization} 
Deciding how often the replicas synchronize
is key to the design. In
Hadoop-based and Bismarck-based models, they synchronize at the end of
each epoch. This is a shared-nothing approach that works well in
user-defined aggregations. However, we consider finer granularities of
sharing.  In \dimm, we chose to have one thread that periodically
reads models on all other \torevise{cores}, averages their results, and updates
each replica. 


\begin{figure}[t]
\centering
\includegraphics[width=0.4\textwidth]{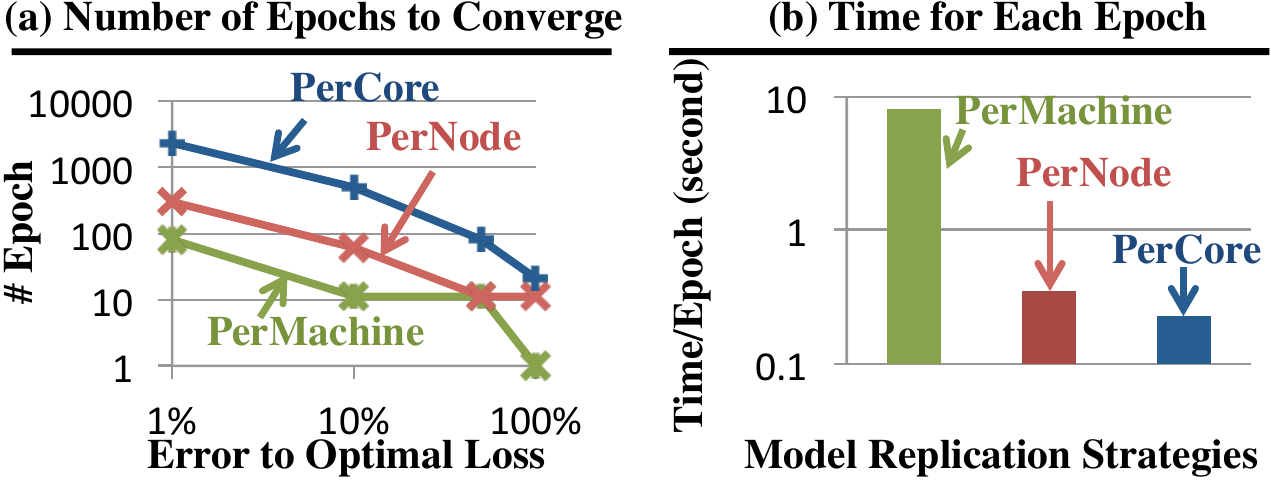}
\caption{Illustration of Model Replication.}
\label{fig:synthetic_model}
\end{figure}

One key question for model synchronization is \yell{{\em how
    frequently should the model be synchronized?}}
Intuitively, we might expect that more frequent synchronization will
lower the throughput; on the other hand, the more frequently we
synchronize, the fewer number of iterations we might need to converge.
However, in \dimm, we find that the optimal choice is to communicate as 
frequently as possible. The intuition
is that the QPI has staggering bandwidth (25GB/s) compared to
the small amount of data we are shipping (megabytes).
As a result, in \dimm, we implement an asynchronous version of the
model averaging protocol: a separate thread averages models,
with the effect of batching many writes together across the
cores into one write, reducing the number of stalls.

\paragraph*{Tradeoffs}
We observe that \pernode is more hardware efficient, as it takes less
time to execute an epoch than \permachine; \permachine might use
fewer number of epochs to converge than \pernode. 

{\bf Statistical Efficiency.} We observe that \permachine usually
takes fewer epochs to converge to the same loss compared to \pernode,
and \pernode uses fewer number of epochs than \percpu. To illustrate
this observation, Figure~\ref{fig:synthetic_model}(a) shows the number
of epochs that each strategy requires to converge to a given loss for
SVM (RCV1). We see that \permachine always uses the least number of
epochs to converge to a given loss: intuitively, the single model
replica has more information at each step, which means \yell{that} 
there is less
redundant work. We observe similar phenomena when comparing \percpu
and \pernode.

{\bf Hardware Efficiency.} We observe that \pernode uses much less
time to execute an epoch than \permachine. To illustrate the difference
\yell{in} the time that each model replication strategy \yell{uses} to finish one
epoch, we show in Figure~\ref{fig:synthetic_model}(b) the execution
time of three strategies on SVM (RCV1). We see that \pernode is 23$\times$
faster than \permachine and \yell{that} \percpu is 1.5$\times$ faster than
\pernode. \pernode takes advantage of the locality provided by the NUMA
architecture. Using PMUs, we find that \permachine incurs 11$\times$ more
cross-node DRAM requests than \pernode.

\paragraph*{Rule of Thumb}
For SGD-based models, \pernode usually gives optimal results\yell{,} while for
SCD-based models, \permachine does.  Intuitively, this is caused by
the fact that SGD has a \yell{denser} update pattern than SCD, \yell{so}, \permachine suffers from hardware efficiency.

\begin{figure}[t]
\centering
\includegraphics[width=0.4\textwidth]{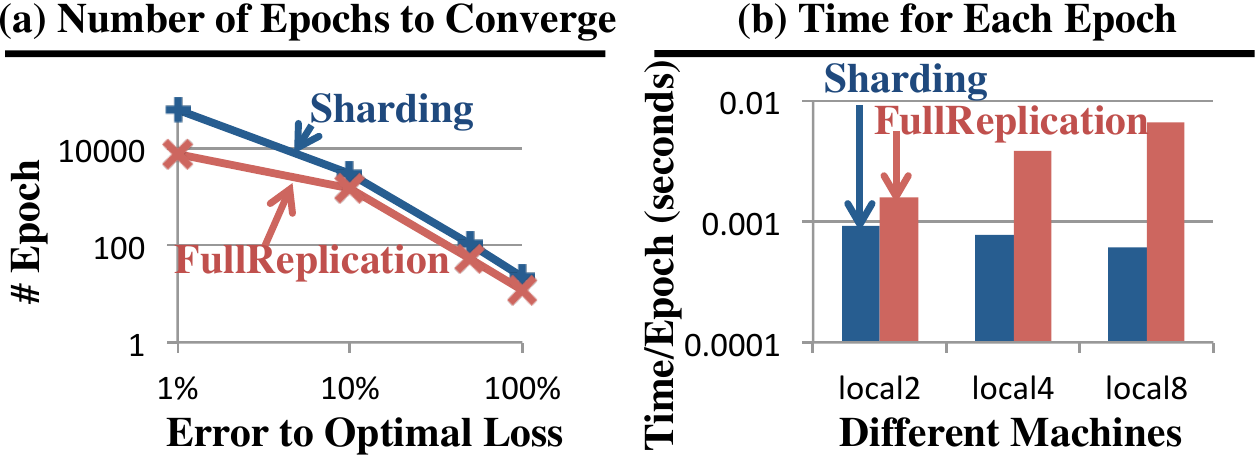}
\caption{Illustration of Data Replication.}
\label{fig:synthetic_data}
\end{figure}

\subsection{Data Replication} \label{sec:data_allocation}

In \dimm, each worker processes a subset of data and then updates its
model replica. To assign a subset of data to each worker, we
consider two strategies.

\paragraph*{\sharding}
\sharding is a popular strategy implemented in systems \yell{such as} \hogwild,
\spark, and Bismarck, in which the dataset is partitioned, and each
worker only works on its partition of data. When there is a single
model replica, \sharding avoids wasted computation\yell{,} as each tuple is
processed once per epoch. However, when there are multiple model
replicas, \sharding might increase the variance of the estimate we form
on each node, lowering the statistical efficiency.
In \dimm, we implement \sharding by randomly partitioning the rows
(resp. columns) of a data matrix for \yell{the} row-wise (resp. column-wise)
access method. In
column-to-row access, we also replicate other rows that
are needed.


\paragraph*{\fullreplication} 
A simple alternative to \sharding is \fullreplication, in which we
replicate the whole dataset many times (\percpu or \pernode). In
\pernode, each NUMA node will have a full copy of the data. Each node
accesses its data in a different order, which means that the replicas
provide non-redundant statistical information. Statistically, there are
two benefits of \fullreplication: (1) averaging different estimates
from each node has a lower variance, and (2) the estimate at each node
has lower variance than in the \sharding case, as each node's estimate
is based on the whole data. From a hardware efficiency perspective,
reads are more frequent from local NUMA memory in \pernode than in
\permachine. The \pernode approach dominates the \percpu approach\yell{,} as
reads from the same node go to the same NUMA memory. Thus, we do not
consider \percpu replication from this point on.

\paragraph*{Tradeoffs}

Not surprisingly, we observe that \fullreplication takes more time for
each epoch than \sharding. However, we also observe that
\fullreplication uses fewer epochs than \sharding,
especially to achieve low error. We illustrate these two observations
by showing the result of running SVM on Reuters using \pernode in
Figure~\ref{fig:synthetic_data}.

{\bf Statistical Efficiency.} 
\fullreplication uses fewer epochs, especially to low-error
tolerance. Figure~\ref{fig:synthetic_data}(a) shows the number of
epochs that each strategy takes to converge to a given loss. We see
that\yell{,} for within 1\% of the loss, \fullreplication uses $10\times$ fewer
epochs on a \yell{two}-node machine. This is because each model
replica sees more data than \sharding, and therefore has a better
estimate. Because of this difference in the number of epochs,
\fullreplication is $5\times$ faster in wall-clock time than \sharding
to converge to 1\% loss. However, we also observe that\yell{,} at high-error
regions, \fullreplication uses more epochs than \sharding and causes a
comparable execution time to a given loss.

{\bf Hardware Efficiency.} 
Figure~\ref{fig:synthetic_data}(b) shows
the time for each epoch across different
machines with different numbers of nodes.
Because we are using the \pernode strategy,
which is the optimal choice for this dataset,
the more nodes a machine has, the slower
\fullreplication is for each epoch.
The slow-down is roughly consistent
with the number of nodes on each machine.
This is not surprising because each epoch
of \fullreplication processes more data than \sharding.\\



\section{Experiments} \label{sec:experiment}

%


\begin{figure}[t]
\centering
\includegraphics[width=0.4\textwidth]{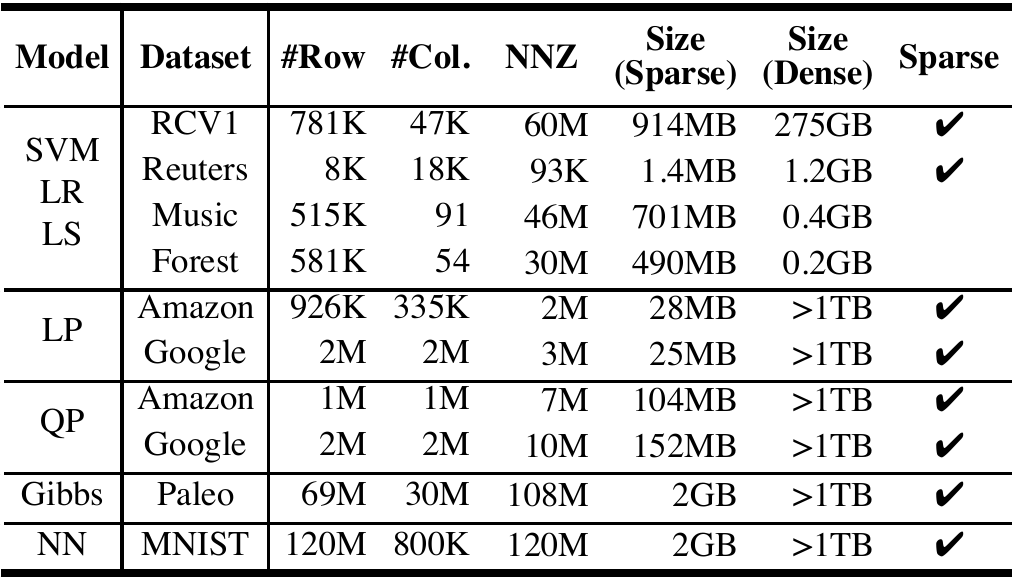}
\caption{Dataset Statistics. 
NNZ refers to the \underline{N}umber of \underline{N}on-\underline{z}ero elements. The \# columns \yell{is} equal to the number of variables in the model.}
\label{tbl:datasets}
\end{figure}


\begin{figure*}[t!]
\scriptsize
\centering
\begin{tabular}{ c c || r | r | r | r | r || r | r | r | r | r  }
\Xhline{4\arrayrulewidth}
\multicolumn{2}{c||}{\multirow{2}{*}{Dataset}}        & \multicolumn{5}{c||}{Within 1\% of the Optimal Loss}             & \multicolumn{5}{c}{Within 50\% of the Optimal Loss} \\
\cline{3-12}
 & & \graphlab & \graphchi & \mllib & \hogwild & DW   & \graphlab & \graphchi & \mllib & \hogwild & DW \\
\hline                       
\multirow{3}{*}{SVM} & Reuters  & $58.9$ & $56.7$ & 15.5 & {\bf \underline{0.1}} & {\bf \underline{0.1}} & $13.6$ & $11.2$ & 0.6 & {\bf \underline{0.01}} & {\bf \underline{0.01}} \\
                     & RCV1 & $>300.0$ & $>300.0$ & $>300$ & 61.4 & {\bf \underline{26.8}}  & $>300.0$ & $>300.0$& $58.0 $ & 0.71 & {\bf \underline{0.17}} \\
                     & Music & $>300.0$ & $>300.0$ & 156 & 33.32 & {\bf \underline{23.7}}  & $31.2$ & $27.1 $& 7.7 & 0.17 & {\bf \underline{0.14}} \\
                     & Forest & $16.2$ & $15.8$ & $ 2.70 $ & 0.23 & {\bf \underline{0.01}}  & $1.9 $ & $1.4$ & $ 0.15 $ & 0.03 & {\bf \underline{0.01}} \\
\hline     
\multirow{3}{*}{LR}  & Reuters & $36.3$ & $34.2$ & $19.2$ & {\bf \underline{0.1}} & {\bf \underline{0.1}} & $13.2$ & $12.5$ & $1.2 $ & {\bf \underline{0.03}} & {\bf \underline{0.03}} \\
                     & RCV1 & $>300.0$ & $>300.0$ & $>300.0$ & 38.7 & {\bf \underline{19.8}}  & $>300.0$ & $>300.0$& $68.0$ & 0.82 & {\bf \underline{0.20}}\\
                     & Music & $>300.0$& $>300.0$ & $>300.0$ & 35.7 &  {\bf \underline{28.6}} & $30.2$ & $28.9$& $8.9$ & 0.56 & {\bf \underline{0.34}}\\
                     & Forest & 29.2 & 28.7 & $3.74 $ & 0.29 & {\bf \underline{0.03}}  & 2.3 & 2.5 & $0.17$ & 0.02 & {\bf \underline{0.01}} \\

\hline  
\multirow{3}{*}{LS}  & Reuters & $132.9$ & $121.2$ & $92.5$ & 4.1 & {\bf \underline{3.2}}   & $16.3$ & $16.7$ & 1.9  & 0.17 & {\bf \underline{0.09}}      \\
                     & RCV1    & $>300.0$ & $>300.0$ & $>300$ & 27.5 & {\bf \underline{10.5}} & $>300.0$ & $>300.0$& $32.0$ & 1.30 & {\bf \underline{0.40}}       \\
                     & Music   & $>300.0$ & $>300.0$ & $221$ & 40.1 & {\bf \underline{25.8}} & $>300.0$ & $>300.0$& $11.2$ & 0.78 & {\bf \underline{0.52}}       \\
                     & Forest  & 25.5 & 26.5 & $1.01$ & 0.33 & {\bf \underline{0.02}} & 2.7 & 2.9 & $0.15$ & 0.04 & {\bf \underline{0.01}}       \\

\hline  
\multirow{2}{*}{LP}  & Amazon & 2.7 & 2.4 & $>120.0$ & $>120.0$ & {\bf \underline{0.94}}  & 2.7 & 2.1 &$120.0$ & 1.86 & {\bf \underline{0.94}} \\
                     & Google & 13.4 & 11.9 & $>120.0$ & $>120.0$ & {\bf \underline{12.56}}  & 2.3 & 2.0& $120.0$ & 3.04 & {\bf \underline{2.02}} \\
\hline  
\multirow{2}{*}{QP}  & Amazon & 6.8 & 5.7 & $>120.0$ & $>120.0$ & {\bf \underline{1.8}}  & 6.8 & 5.7 & $>120.0$ & $>120.00$ & {\bf \underline{1.50}}\\
                     & Google & 12.4 & 10.1 & $>120.0$ & $>120.0$ & {\bf \underline{4.3}}  & 9.9 & 8.3 & $>120.0$ & $>120.00$ & {\bf \underline{3.70}}\\
\Xhline{4\arrayrulewidth}
\end{tabular}
\caption{End-to-End Comparison (time in seconds). 
\torevise{The column DW refers to \dimm.}
We take $5$ runs
on local2 and report the average (standard deviation for all 
numbers $< 5\%$ of the mean).
Entries with $>$ indicate a timeout. }
\label{tbl:main}
\end{figure*}

We validate that exploiting the tradeoff space 
that we described enables \dimm's orders of magnitude speedup
over state-of-the-art competitor systems.
We also validate that each tradeoff discussed
in this paper affects the performance of \dimm.

\subsection{Experiment Setup}

We describe the details of our experimental setting.

\paragraph*{Datasets and Statistical Models}

\begin{sloppypar}
We validate the performance and quality of \dimm on a diverse set of
statistical models and datasets. For statistical models, we choose
five models that are among the most popular models used in statistical
analytics: (1) Support Vector Machine (SVM), (2) Logistic Regression
(LR), (3) Least Squares Regression (LS), (4) Linear Programming (LP),
and (5) Quadratic Programming (QP).
For each model, we choose datasets
with different characteristics, including
size, sparsity, and under- or over-determination.
For SVM, LR, and LS, we choose four datasets:
Reuters\footnote{\url{archive.ics.uci.edu/ml/datasets/Reuters-21578+Text+Categorization+Collection}},
RCV1\footnote{\url{about.reuters.com/researchandstandards/corpus/}},
Music\footnote{\url{archive.ics.uci.edu/ml/datasets/YearPredictionMSD}},
and Forest.\footnote{\url{archive.ics.uci.edu/ml/datasets/Covertype}}
Reuters and RCV1 are datasets for
text classification that are sparse
and underdetermined.
Music and Forest are standard benchmark datasets that are dense and
overdetermined.
For QP and LR, we consider a social-network application, i.e., network
analysis, and use two datasets from Amazon's customer data and
Google's Google+ social
networks.\footnote{\url{snap.stanford.edu/data/}}
Figure~\ref{tbl:datasets} shows 
the dataset statistics.
\end{sloppypar}

\begin{sloppypar}
\paragraph*{Metrics} We measure the quality and performance of \dimm and
other competitors.
To measure the quality, we follow prior art and use the loss function
for all functions.
For end-to-end performance, we measure the wall-clock time it takes
for each system to converge to a loss that is within 100\%, 50\%,
10\%, and 1\% of the optimal loss.\footnote{We \yell{obtain} the optimal loss by
  running all systems for one hour and \yell{choosing} the lowest.}
When measuring the wall-clock time, we do not count the time
used for data loading and result outputting for all systems.
We also use other measurements to understand the details of
the tradeoff space,
including (1) \yell{local} LLC request, (2) \yell{remote} LLC request, and
(3) \yell{local} DRAM request.
We use Intel Performance Monitoring Units (PMUs)
and follow \yell{the} manual\footnote{\url{software.intel.com/en-us/articles/performance-monitoring-unit-guidelines}} to conduct these 
experiments. 
\end{sloppypar}

\begin{figure*}[t!]
\centering
\includegraphics[width=0.8\textwidth]{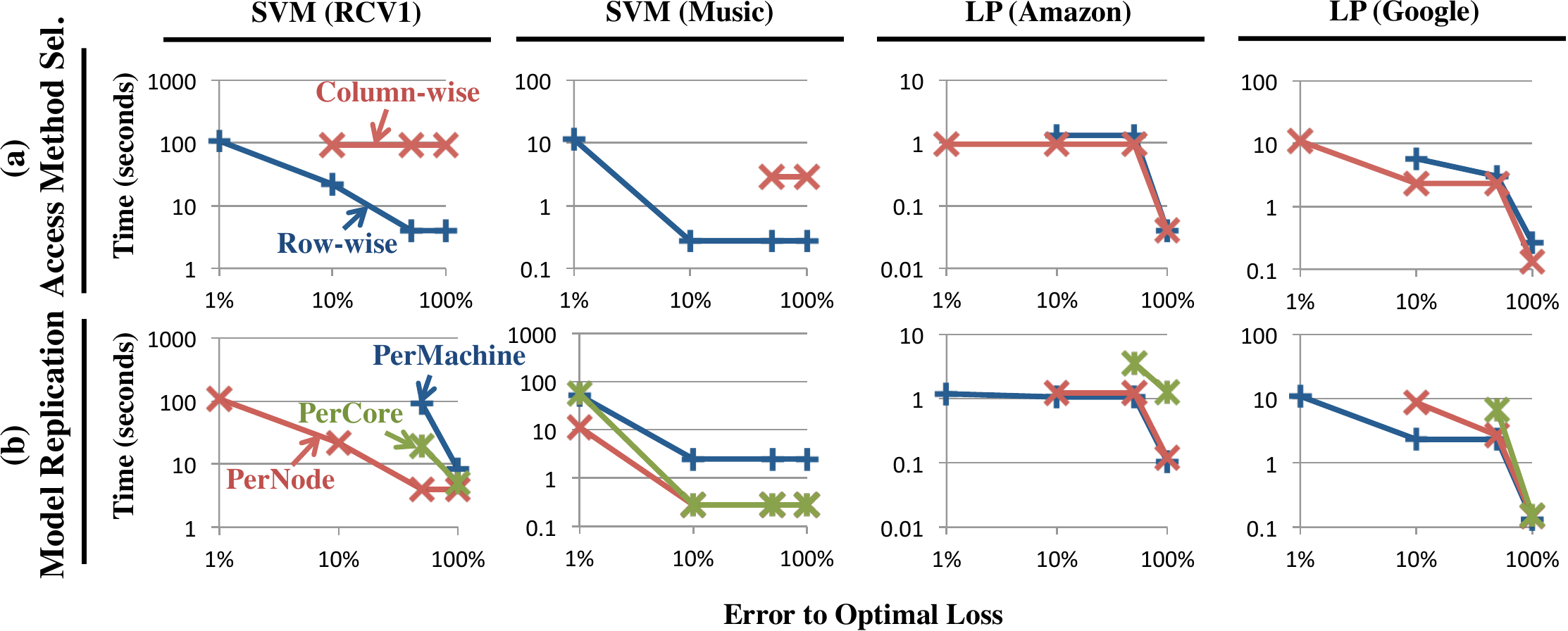}
\caption{Tradeoffs in \dimm.
Missing points timeout in 120 seconds.}
\label{fig:data_algo_lesion}
\end{figure*}

\paragraph*{Experiment Setting}

We compare \dimm with \torevise{four} competitor systems:
\graphlab~\cite{DBLP:journals/pvldb/LowGKBGH12},
\graphchi~\cite{Kyrola:2012:OSDI},
\torevise{\mllib~\cite{Sparks:2013:ICDM} over \spark~\cite{Zaharia:2012:NSDI}}, and
\hogwild~\cite{DBLP:conf/nips/RechtRWN11}.
\graphlab is a distributed graph processing
system that supports a large range of statistical
models. \torevise{\graphchi is similar to \graphlab
but with \yell{a} focus on multi-core machines with secondary
storage.}
\torevise{
\mllib is a package of machine learning algorithms
implemented over \spark, an in-memory implementation of
the MapReduce framework.
}
%
%
\hogwild is an in-memory lock-free framework
for statistical analytics.
We find that all \torevise{four} systems pick some points in the tradeoff space
that we considered in \dimm. In \graphlab and \graphchi, all models are implemented
using stochastic coordinate descent (column-wise access); in
\torevise{\mllib} and \hogwild, SVM and LR are implemented using
stochastic gradient descent (row-wise access). We use implementations
that are provided by the original developers whenever possible.  For
models without code provided by the developers, we only change the
corresponding gradient function.\footnote{\torevise{For sparse models,
we change the dense vector data structure in \mllib to a sparse vector,
which only improves its performance.}} \torevise{For \graphchi, if the corresponding
model is implemented in \graphlab but not \graphchi, we follow
\graphlab's implementation.}

We run experiments on a variety of architectures. These machines
differ in a range of configurations, including \yell{the} 
number of NUMA nodes,
the size of last-level cache (LLC), and memory bandwidth. See
Figure~\ref{fig:architecture} for a summary of these machines.
\dimm, \hogwild, \graphlab, and \torevise{\graphchi} are implemented using C++\yell{,}
and \torevise{\mllib/\spark} is implemented using Scala.
We tune both \graphlab and \torevise{\mllib} according to their best
practice
guidelines.\footnote{\torevise{\mllib}:\url{spark.incubator.apache.org/docs/0.6.0/tuning.html};
  \graphlab:
  \url{graphlab.org/tutorials-2/fine-tuning-graphlab-performance/}.
  \torevise{ For \graphchi, we tune the memory buffer size to \yell{ensure} all data fit in memory and \yell{that} there are no disk I/Os.  We
    describe more detailed tuning for \torevise{\mllib} in the full
    version of this paper.}  } \torevise{For both \graphlab,
  \graphchi, and \mllib, we try different ways of increasing locality
  on NUMA machines, including trying to use \textsf{numactl} and
  \yell{implementing} our own RDD for \mllib\yell{;} there is more detail in the full
  version of this paper.}
Systems are compiled with g++ 4.7.2 (-O3), Java 1.7, or Scala 2.9.

\subsection{End-to-End Comparison}

We validate that \dimm outperforms competitor systems in terms of
end-to-end performance and quality.  Note that both \mllib and
\graphlab have extra overhead for fault tolerance, distributing work,
and task scheduling. Our comparison between \dimm and these
competitors is intended only to demonstrate that existing work for
statistical analytics has not obviated the tradeoffs that we study
here. 

\begin{figure}[t!]
\centering
\includegraphics[width=0.4\textwidth]{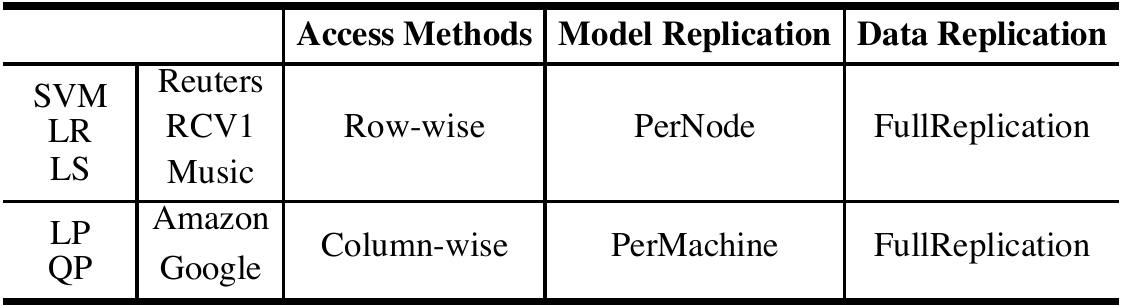}
\caption{Plans that \dimm Chooses in the Tradeoff Space for Each Dataset on Machine local2.}
\label{fig:plan_of_choice}
\end{figure}

\paragraph*{Protocol}
For each system, we grid search their statistical parameters,
including step size (\{100.0,10.0,...,0.0001\}) and mini-batch size
for \torevise{\mllib} (\{1\%, 10\%, 50\%, 100\%\}); we always report
the best configuration, which is essentially the same for each
system. We measure the time it takes for each system to find a
solution that is within 1\%, 10\%, and 50\% of the optimal
loss. Figure~\ref{tbl:main} shows the result\yell{s} for 1\% and 50\%; the
results for 10\% are similar. We report end-to-end numbers from local2\yell{, which has two}
nodes and 24 logical cores, as \graphlab does not run on
machines with more than 64 logical
cores. Figure~\ref{fig:plan_of_choice} shows the \dimm's choice of
point in the tradeoff space on local2.

As shown in Figure~\ref{tbl:main}, \dimm always converges to the given
loss in less time than the other competitors. On SVM and LR, \dimm
could be up to 10$\times$ faster than \hogwild, and more than two
orders of magnitude faster than \graphlab and \spark.
The difference between \dimm and \hogwild is greater
for LP and QP, where \dimm outperforms \hogwild by
more than two orders of magnitude.
On LP and QP, \dimm is also up to 3$\times$ faster than \graphlab
and \torevise{\graphchi}, and two orders of magnitude faster than \torevise{\mllib}.

\paragraph*{Tradeoff Choices}
We dive more deeply into these numbers to substantiate our claim that
there are some points in the tradeoff space that are not used by
\graphlab, \torevise{\graphchi}, \hogwild, and \torevise{\mllib}. Each
tradeoff selected by our system is \yell{shown} in
Figure~\ref{fig:plan_of_choice}. For example, \graphlab and \graphchi
uses column-wise access for all models, while \torevise{\mllib} and
\hogwild use row-wise access for all models and allow only \permachine
model replication. These special points work well for some \yell{but not
all} models.  For example, for LP and QP, \graphlab and
  \graphchi are only 3$\times$ slower than \dimm, which chooses
column-wise and \permachine. This factor of 3 is to be expected\yell{,} as
GraphLab also allows distributed access and so has additional
overhead. \yell{However} there are other points: \yell{for} SVM and LR, \dimm outperforms
\graphlab and \graphchi, because the column-wise algorithm
implemented by \graphlab and \graphchi is not as efficient
as row-wise on the same dataset. \dimm outperforms \hogwild because
\dimm takes advantage of model replication, while \hogwild incurs
11$\times$ more cross-node DRAM requests than \dimm; in contrast,
\dimm incurs 11$\times$ more local DRAM requests than \hogwild does.

For SVM, LR, and LS, we find that \dimm outperforms
  \mllib\yell{,} primarily due to a different point in the tradeoff
  space. In particular, \mllib uses batch-gradient-descent with a
  \percpu implementation, while \dimm uses stochastic gradient and
  \pernode.  We find that\yell{, for} the Forest dataset\yell{,} \dimm takes
  60$\times$ fewer number of epochs to converge to 1\% loss than
  \torevise{\mllib}. For each epoch, \dimm is 4$\times$ faster. These
  two factors contribute to the 240$\times$ speed-up of \dimm over
  \mllib on the Forest dataset (1\% loss).  \mllib has overhead for
  scheduling, so we break down the time that \mllib \yell{uses} for
  scheduling and computation. We find that\yell{,} for Forest, out of the
  total 2.7 seconds of execution, \mllib uses 1.8 seconds for
  computation and 0.9 seconds for scheduling. We also implemented a
  batch-gradient-descent and \percpu implementation inside \dimm to
  remove these and C++ versus Scala differences. The 60$\times$
  difference in \yell{the} number of epochs until convergence still holds, and
  our implementation is only 3$\times$ faster than \mllib. This
  implies that the main difference between \dimm and \mllib is the
  point in the tradeoff space---not low-level implementation differences. 

\yell{For} LP and QP, \dimm outperforms \torevise{\mllib} and
  \hogwild because the row-wise access method implemented by these
  systems \yell{is} not as efficient as column-wise access on the same data
  set. \graphlab does have column-wise access, so \dimm
  outperforms \graphlab and \graphchi because \dimm
  finishes each epoch up to 3$\times$ faster\yell{,} primarily due to
  low-level issues. This supports our claims that the tradeoff space
  is interesting for analytic engines and \yell{that} no one system has
  implemented all of them.


\begin{figure}[t!]
\centering
\includegraphics[width=0.4\textwidth]{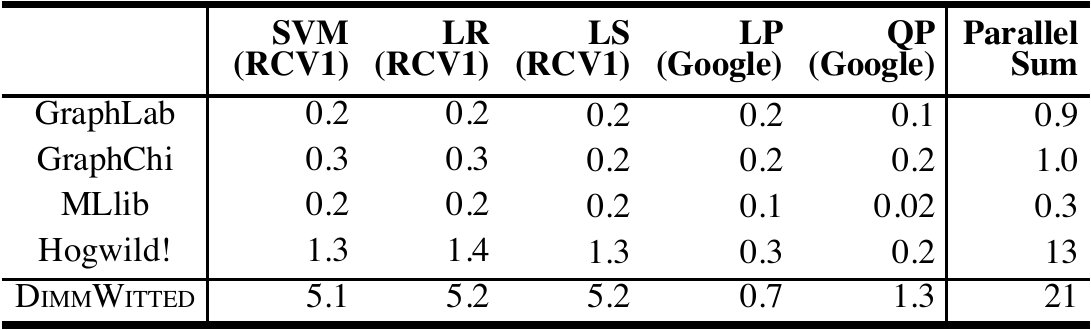}
\caption{Comparison of Throughput (GB/seconds) of Different Systems on local2.}
\label{tbl:throughput}
\end{figure}

\paragraph*{Throughput}
We compare the throughput of different systems for an extremely simple
task: parallel sums.
Our implementation of parallel sum follows our implementation of other
statistical models (with a trivial update function), and uses all
cores on a single machine.
Figure~\ref{tbl:throughput} shows the throughput on 
all systems on different models on one dataset.
We see from Figure~\ref{tbl:throughput} that \dimm
achieves the highest throughput of all the systems.
For parallel sum, \dimm is 1.6$\times$ faster 
than \hogwild, and we find that
\dimm incurs $8\times$ fewer LLC cache
misses than \hogwild.
Compared with \hogwild, in which all threads write to a single copy of
the sum result, \dimm maintains one single copy of the sum result per
NUMA node, \yell{so} the workers on one NUMA node do not
invalidate the cache on another NUMA node.
When running on only a single thread, \dimm has the same
implementation as \hogwild.
Compared with \graphlab and \graphchi, \dimm is $20\times$ faster,
likely due to the overhead of \graphlab and \graphchi
dynamically scheduling tasks and/or maintain\yell{ing} the graph structure. 
To compare \yell{\dimm} with \torevise{\mllib}, which is written in Scala, we implemented a
Scala version, which is $3\times$ slower than C++; this
suggests that the overhead is not just due to the language.
If we do not count the time that \mllib \yell{uses} for scheduling and
only count the time of computation, we find that \dimm is
15$\times$ faster than \torevise{\mllib}.


\subsection{Tradeoffs of \dimm}

We validate that all \yell{the} tradeoffs described in this paper have an impact
on the efficiency of \dimm. We report on a more modern architecture,
local4 with \yell{four} NUMA sockets, in this section. We describe how the
results change with different architecture\yell{s}.

\begin{figure}[t!]
\centering
\includegraphics[width=0.4\textwidth]{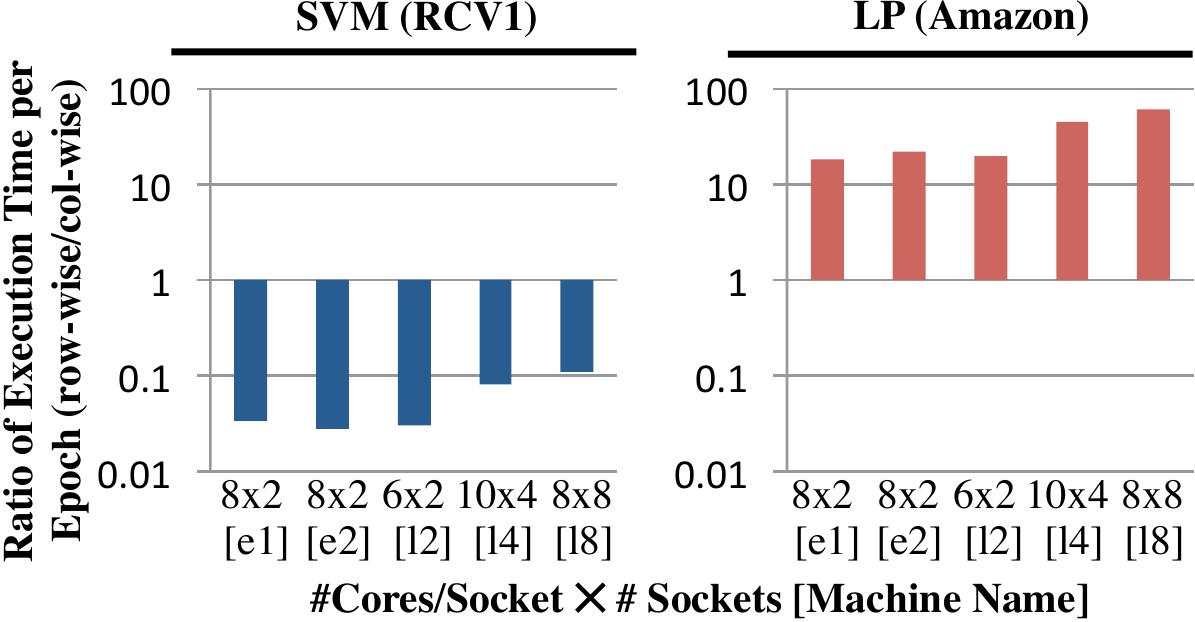}
\caption{Ratio of Execution Time per Epoch (row-wise/column-wise) on Different Architectures. A number larger than 1 means \yell{that} row-wise is slower. l2 means local2, e1 means ec2.1, etc.}
\label{fig:row_vs_column}
\end{figure}

\subsubsection{Access Method Selection}
We validate that different access methods have different performance,
and that no single access method dominates the others.  We run
\dimm on all statistical models and compare two strategies, row-wise
and column-wise. In each experiment, we force \dimm to use the
corresponding access method, but report the best point for the other
tradeoffs. Figure~\ref{fig:data_algo_lesion}(a) shows the results as
we measure the time it takes to achieve each loss. The more stringent
loss requirements ($1\%$) are on the left-hand side. The horizontal
line segments in the graph indicate that a model may reach, say,
$50\%$ as quickly (in epochs) as it reaches $100\%$.

We see from Figure~\ref{fig:data_algo_lesion}(a) that the difference
between row-wise and column-to-row access could be more than
$100\times$ for different models. For SVM on RCV1, row-wise access
converges at least 4$\times$ faster to 10\% loss and at least
$10\times$ faster to 100\% loss. We observe similar phenomena
\yell{for} Music; compared with RCV1, column-to-row access converges to 50\%
loss and 100\% loss at a 10$\times$ slower rate. \yell{With} such
datasets, the column-to-row access simply requires more reads and
writes. This supports the folk wisdom that gradient methods are
preferable to coordinate descent methods. On the other hand,
for LP, column-wise access dominates: row-wise access does not
converge to 1\% loss within the timeout period for either Amazon or
Google. Column-wise access converges at least 10-100$\times$
faster than row-wise access to 1\% loss.  We observe that LR is
similar to SVM and QP is similar to LP. Thus, no access method
dominates all \yell{the} others.

The cost of writing and reading are different and is captured by a
parameter that we called $\alpha$ in
Section~\ref{sec:method_selection}. We describe the impact of this factor
on the relative performance of row- and column-wise
strategies. Figure~\ref{fig:row_vs_column} shows the ratio of the time
that each strategy \yell{uses} (row-wise/column-wise) \yell{for} SVM (RCV1) and LP (Amazon).
We see that, as the number of sockets on a machine increases, the
ratio of execution time \yell{increases}, which means that row-wise \yell{becomes}
slower relative to column-wise, i.e., with increasing $\alpha$. As the
write cost captures the cost of a hardware-resolved conflict, we see
that this constant is likely to grow. Thus, if \yell{next-generation}
architectures increase in \yell{the} number of sockets, the cost parameter 
$\alpha$ and consequently the importance of this tradeoff are likely to grow.

\paragraph*{Cost-based Optimizer} 
We observed that\yell{, for} all datasets, our cost-based optimizer selects
row-wise access for SVM, LR, and LS, and column-wise access for LP and
QP. These choices are consistent with what we observed in
Figure~\ref{fig:data_algo_lesion}.

\subsubsection{Model Replication}

We validate that there is no single strategy for model replication
that dominates the others. We force \dimm to run strategies in
\permachine, \pernode, and \percpu and choose other tradeoffs by
choosing the plan that achieves the best
result. Figure~\ref{fig:data_algo_lesion}(b) shows the \yell{results}.

We see from Figure~\ref{fig:data_algo_lesion}(b) that the gap between
\permachine and \pernode could be up to 100$\times$. We first observe
that \pernode dominates \percpu on all datasets. For SVM on RCV1,
\pernode converges 10$\times$ faster than \percpu to 50\% loss, and \yell{for}
other models and datasets, we observe a similar phenomenon. This is due
to the low statistical efficiency of \percpu, as we discussed in
Section~\ref{sec:model_allocation}. Although \percpu eliminates write
contention inside one NUMA node, this write contention is less
critical. \yell{For} large models and machines with small caches, we also
observe that \percpu could spill the cache.

These graphs show that neither \permachine nor \pernode dominates the
other across all datasets and statistical models. For SVM on RCV1,
\pernode converges 12$\times$ faster than \permachine to 50\%
loss. However, for LP on Amazon, \permachine is at least 14$\times$
faster than \pernode to converge to 1\% loss. For SVM,
\pernode converges faster because it has 5$\times$ higher
throughput than \permachine, and for LP, \pernode is
slower because \permachine takes at least 10$\times$ fewer
epochs to converge to a small loss. One interesting observation is
that\yell{,} for LP on Amazon, \permachine and \pernode do have comparable
performance to converge to 10\% loss. Compared with the 1\% loss case,
this implies that \pernode's statistical efficiency decreases as the
algorithm tries to achieve a smaller loss. This is not surprising, as
one must reconcile the \pernode estimates.

We observe that the relative performance of \permachine and \pernode
depends on (1) the number of sockets used on each machine and (2) the
sparsity of the update.

\begin{figure}[t!]
\centering
\includegraphics[width=0.45\textwidth]{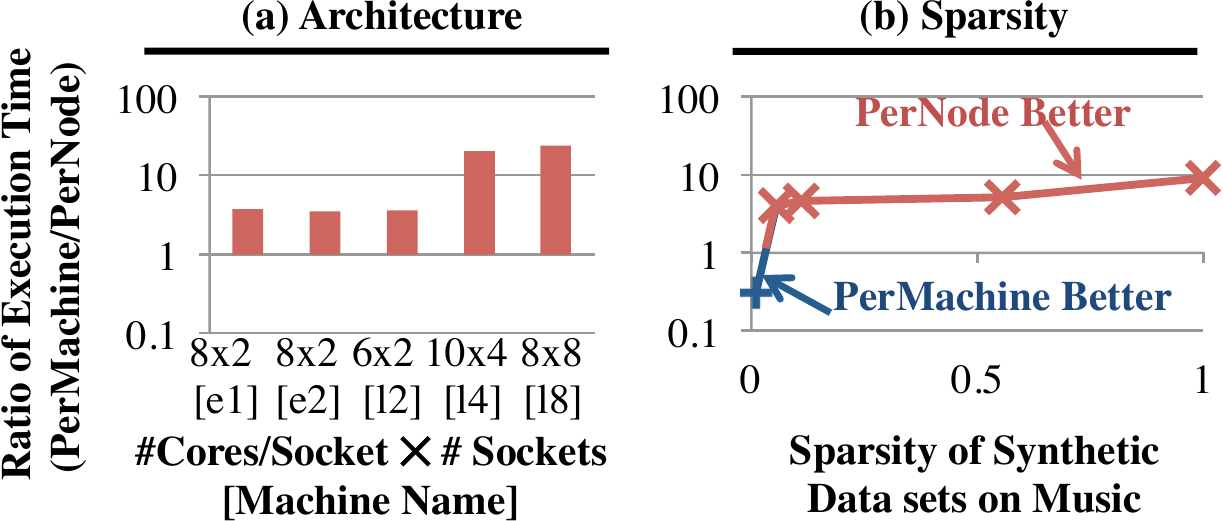}
\caption{The Impact of Different Architectures and Sparsity on Model Replication. A
ratio larger than 1 means that \pernode converges faster than \permachine to 50\% loss.}
 \label{fig:result_model_replication}
 \end{figure}

\begin{sloppypar}
To validate (1), we measure the time that \pernode and
\permachine take on SVM (RCV1) to converge to 50\% loss on various
architectures, and we report the ratio (\permachine/\pernode) in
Figure~\ref{fig:result_model_replication}. We see that \pernode's
relative performance improves with the number of sockets. We attribute
this to the increased cost of write contention in \permachine.
\end{sloppypar}

To validate (2), we generate a series of synthetic datasets\yell{,} each of
which subsamples the elements in each row of the Music dataset;
Figure~\ref{fig:result_model_replication}(b) shows the \yell{results}.  When
the sparsity is 1\%, \permachine outperforms \pernode, as each update
touches only one element of the model; thus, the write contention in
\permachine is not a bottleneck. As the sparsity increases (i.e.,
the update \yell{becomes denser}), we observe that \pernode outperforms
\permachine.

\begin{figure}[t!]
\centering
\includegraphics[width=0.45\textwidth]{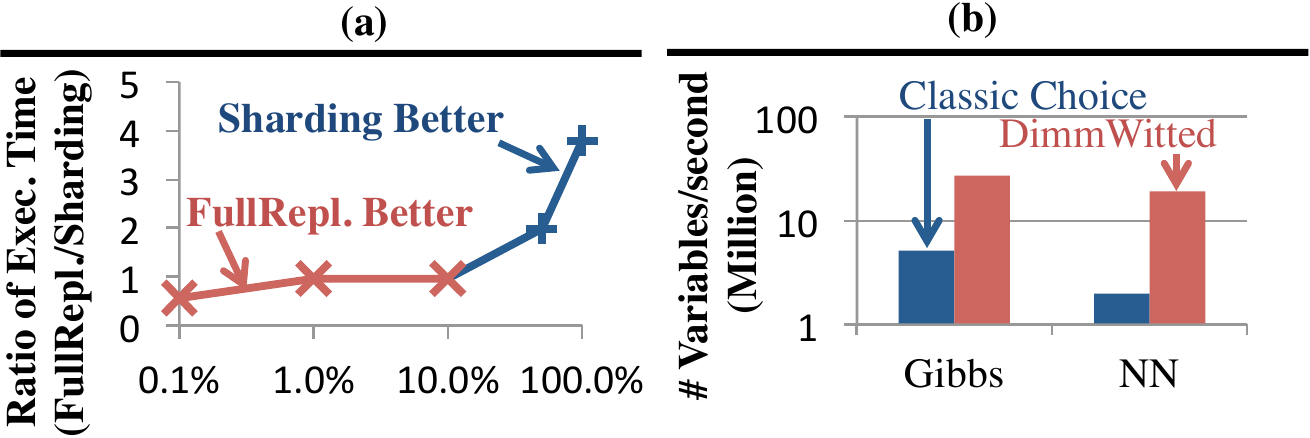}
\caption{(a) Tradeoffs of Data Replication. A ratio smaller than 1 means that
\fullreplication is faster. (b) Performance of Gibbs Sampling and Neural Networks
Implemented in \dimm.}
\label{fig:result_data_replication}
\end{figure}

\subsubsection{Data Replication}

We validate the impact of different data replication strategies. We
run \dimm by fixing data replication strategies to \fullreplication or
\sharding and \yell{choosing} the best plan for each other tradeoff.  We
measure the execution time for each strategy to converge to a
given loss for SVM on the same dataset, RCV1. We report the ratio of
these two strategies as \fullreplication/\sharding in
Figure~\ref{fig:result_data_replication}(a).
We see that\yell{,} for \yell{the} low-error region (e.g., 0.1\%),
\fullreplication is 1.8-2.5$\times$
faster than \sharding.
This is because \fullreplication decreases the skew of data assignment
to each worker, \yell{so} hence each individual model replica can form a
more accurate estimate.
For the high-error region (e.g., 100\%), we observe that
\fullreplication appears to be 2-5$\times$ slower than
\sharding.
We find that, for 100\% loss, both \fullreplication
and \sharding converge in a single epoch\yell{,} and \sharding may therefore be
preferred, as it examines less data to complete that single epoch.
In all of our experiments, \fullreplication is never substantially
worse and can be dramatically better. Thus, if there is available
memory, the \fullreplication data replication seems to be preferable.

\section{Extensions} \label{sec:extension}



We briefly describe how to run Gibbs sampling (which uses a
column-to-row access method) and deep neural networks (which uses a
row access method).
Using the same tradeoffs, we achieve \yell{a} significant increase in speed
over \yell{the} classical implementation choices of these algorithms.
\torevise{A more detailed description is in the full version
  of this paper.}

\subsection{Gibbs Sampling}

Gibbs sampling is one of the most popular algorithms to solve
statistical inference and learning over probabilistic graphical
models~\cite{Robert:2005:Book}. We briefly describe Gibbs sampling
over factor graphs and observe \yell{that} its main step is a column-to-row access.
A factor graph can be thought of as a bipartite graph of a set
of variables and a set of factors. To run Gibbs sampling, the main
operation is to select a single variable, and calculate the
conditional probability of this variable, which requires \yell{the fetching of} all
factors that contain this variable and all assignments of variables
connected to these factors. This operation corresponds to the
column-to-row access method. Similar to \yell{first-order} methods, recently\yell{,} a Hogwild! algorithm for Gibbs was
established~\cite{Johnson:2013:NIPS}. As shown in
Figure~\ref{fig:result_data_replication}(b), applying the technique in
\dimm to Gibbs sampling \yell{achieves} 4$\times$ the throughput of samples
as the \permachine strategy. 

\subsection{Deep Neural Networks}

Neural networks are one of the most classic machine learning
models~\cite{Mitchell:1997:Book}; recently\yell{,} these models have
  been intensively revisited by adding more
layers~\cite{DBLP:conf/icml/LeRMDCCDN12,DBLP:conf/nips/DeanCMCDLMRSTYN12}.
A deep neural network contains multiple layers\yell{, and} each layer
contains a set of neurons (variables). Different neurons connect with
each other only by links across consecutive layers.  The value of one
neuron is a function of all \yell{the} other neurons in the previous layer and a
set of weights.  Variables in the last layer have human labels as
training data; the goal of deep neural network learning is to find the
set of weights that maximizes the likelihood of the human
labels. Back-propagation with stochastic gradient descent is the de
facto method of optimizing a deep neural network.

Following LeCun et al.~\cite{LeCun:1998:PIEEE}, we implement SGD over
a seven-layer neural network with 0.12 billion neurons and 0.8 million
parameters using a standard handwriting-recognition benchmark dataset
called MNIST\footnote{\url{yann.lecun.com/exdb/mnist/}}.
Figure~\ref{fig:result_data_replication}(b) shows the number of
variables that are processed by \dimm per second. For this
application, \dimm uses \pernode and \fullreplication, and the
classical choice made by LeCun is \permachine and \sharding. As
  shown in Figure~\ref{fig:result_data_replication}(b), \dimm
achieves more than an order of magnitude higher throughput than this
classical baseline (to achieve the same quality as reported in this
classical paper).

\section{Related Work}
\label{sec:related:work}
We review work in \torevise{four} main areas: statistical analytics, data mining
algorithms, \torevise{shared-memory multiprocessors optimization}, and main-memory 
databases. We include more extensive
related work in the full version \yell{of this paper}.

\textbf{Statistical Analytics.}  There is a trend to integrate
statistical analytics into data processing systems. Database vendors
have recently put out new products in this space, including Oracle,
Pivotal's MADlib~\cite{Hellerstein:2012:VLDB}, IBM's
SystemML~\cite{Ghoting:2011:ICDE}, and SAP's HANA. These systems
support statistical analytics in existing data management systems. A
key challenge for statistical analytics is performance.

\begin{sloppypar}
A handful of data processing frameworks have been developed in the
last few years to support statistical analytics, including Mahout for
Hadoop, MLI for \spark~\cite{Sparks:2013:ICDM},
\graphlab~\cite{DBLP:journals/pvldb/LowGKBGH12}, and MADLib for
PostgreSQL or Greenplum~\cite{Hellerstein:2012:VLDB}. Although these
systems increase the performance of corresponding statistical
analytics tasks significantly, we observe that each of them
implements one point in \dimm's tradeoff space. \dimm is not a system;
our goal is to study this tradeoff space.
\end{sloppypar}

\textbf{Data Mining Algorithms.}  There is a large \yell{body of}
data mining literature regarding how to optimize various algorithms to
be more architecturally
aware~\cite{Parthasarathy:2001:KAIS,Zaki:1997:KDD,Zaki:1999:ICDE}.
Zaki et al.~\cite{Parthasarathy:2001:KAIS,Zaki:1997:KDD} \yell{study} the
performance of a range of different algorithms, including associated
rule mining and decision tree on shared-memory machines, by 
\yell{improving} memory
locality \yell{and} data placement in the granularity of cachelines, and
decreasing \yell{the} cost of coherent maintenance between multiple CPU
caches. Ghoting et al.~\cite{Ghoting:2007:VLDBJ} optimize
\yell{the} cache behavior of frequent pattern mining using novel cache-conscious
techniques, including spatial and temporal locality, prefetching, and
tiling. Jin et al.~\cite{Jin:2005:TKDE} discuss tradeoffs in
replication and locking schemes for K-means, association rule mining,
and neural nets. This work considers the hardware efficiency of the
algorithm, but not statistical efficiency, which is the focus of
\dimm.  In addition, Jin et al. \yell{do} not consider lock-free execution,
a key aspect of this paper.

\textbf{Shared-memory Multiprocessor Optimization.} 
Performance optimization on shared-memory multiprocessors machines is
a classical topic. Anderson and Lam~\cite{Anderson:1993:PLDI} and Carr
et al.'s~\cite{Carr:1994:ASPLOS} seminal work used complier techniques
to improve locality on shared-memory multiprocessor machines. \dimm's
{\em locality group} is inspired by Anderson and Lam's discussion of
{\em computation decomposition} and {\em data decomposition}. These
locality groups \yell{are} the centerpiece of the Legion
project~\cite{DBLP:conf/sc/BauerTSA12}.
\torevise{ In recent years, there have been a variety of {\em domain
    specific languages} (DSLs) to help the user extract parallelism;
  \yell{two} examples of these DSLs include
  Galois~\cite{Nguyen:2014:ASPLOS,Nguyen:2013:SOSP} and
  OptiML~\cite{Sujeeth:2011:ICML} for Delite~\cite{Chafi:2011:PPOPP}.
Our goals are orthogonal: these DSLs require knowledge about the
trade-offs of the hardware, such as \yell{those} provided by our study.  }

\textbf{Main-memory Databases.} The database community has recognized
 that multi-socket, large-memory machines have changed the data
 processing landscape, and there \yell{has been} a flurry of recent work about how
 to build in-memory analytics
 systems~\cite{Albutiu:2012:VLDB,Raman:2013:VLDB,
   Balkesen:2013:VLDB,Tu:2013:SOSP,Li:2013:CIDR,Qiao:2008:VLDB,
   Kim:2009:VLDB,Chasseur:2013:VLDB}.
Classical tradeoffs have been revisited on
modern architectures to gain significant improvement:
Balkesen et al.~\cite{Balkesen:2013:VLDB},
Albutiu et al.~\cite{Albutiu:2012:VLDB},
Kim et al.~\cite{Kim:2009:VLDB},
and Li~\cite{Li:2013:CIDR} \yell{study} the tradeoff
for joins and shuffling, respectively.
This work takes advantage of modern architectures, e.g., NUMA and
SIMD, to increase memory bandwidth. We study a new tradeoff space for
statistical analytics in which the performance of the system is
affected by both hardware efficiency and statistical efficiency.

\section{Conclusion}

For statistical analytics on main-memory, NUMA-aware machines, we
studied tradeoffs in access methods, model replication, and data
replication. We found that using novel points in this tradeoff space
can have a substantial benefit: our \dimm prototype engine can run at
least one popular task at least 100$\times$ faster than other
competitor systems. This comparison demonstrates that this tradeoff
space may be interesting for current and \yell{next-generation} 
statistical analytics systems. \\

\noindent
{\scriptsize{\textbf{Acknowledgments} We would like to thank Arun Kumar, Victor
    Bittorf, the Delite team, the Advanced Analytics at Oracle,
    Greenplum/Pivotal, and Impala's Cloudera team for sharing their
    experiences in building analytics systems.
We gratefully acknowledge the support of the
Defense Advanced Research Projects Agency
(DARPA) XDATA Program under No. FA8750-12-2-0335
and the DEFT Program under No. FA8750-13-2-0039, the National Science
Foundation (NSF)
CAREER Award under No. IIS-1353606, 
the Office of Naval Research (ONR) under awards 
No. N000141210041 and No. N000141310129,
the Sloan Research Fellowship, American Family 
Insurance, Google, and Toshiba.
Any opinions, findings, and conclusion or recommendations
expressed in this material are
those of the authors and do not necessarily reflect
the view of DARPA, NSF, ONR, or the
US government.
}
}

{\scriptsize \bibliographystyle{abbrv} \bibliography{dimm}}

\newpage
\appendix

\section{Implementation Details} \label{sec:impl_details}

In \dimm, we implement optimizations that are part of scientific
computation and analytics systems.  While these optimizations are not
new, they are not universally implemented in analytics systems. We
briefly describes each optimization and its impact.

\paragraph*{Data and Worker Collocation}

We observe that different strategies of locating data and workers
affect the performance of \dimm.
One standard technique is to \yell{collocate}
the worker \yell{and} the data on the same NUMA node.
In this way, \yell{the worker} in each node will pull 
data from its own DRAM region, and does not
need to occupy the node-DRAM bandwidth
of other nodes.
In \dimm, we tried two different placement strategies for 
data and workers.
The first protocol, called \OS, relies on the operating system to
allocate data and threads for workers.
The operating system will usually locate
data on one single NUMA node, and worker threads
to different NUMA \yell{nodes} using heuristics
that are not exposed to the user.
The second protocol, called \NUMA, evenly \yell{distributes} worker threads \yell{across}
NUMA \yell{nodes}, and for each worker, \yell{replicates} the data on the same NUMA
node. We find that for SVM on RCV1, the strategy \NUMA can be up to
2$\times$ faster than \OS. \yell{Here are two reasons for this} improvement. 
First, by locating data on the same NUMA node to workers, we
achieve 1.24$\times$ improvement on the throughput of reading
data. Second, by not asking the operating system to allocate workers,
we actually have a more balanced allocation of workers on NUMA nodes.

\paragraph*{Dense and Sparse} \label{sec:dense_sparse}

For statistical analytics workloads, it is not
uncommon for the data matrix $A$ to be sparse,
especially for applications such as information extraction
and text mining.
In \dimm, we implement two protocols, \Dense
and \Sparse, which \yell{store} the data matrix
$A$ as a dense or sparse matrix, respectively.
A \Dense storage format has two advantages: (1) if 
storing a fully dense vector, it requires $\frac{1}{2}$ the 
space as a sparse representation, 
and (2) \Dense is able to leverage hardware SIMD 
instructions, which allows multiple floating point
operations to be \yell{performed} in parallel.
A \Sparse storage format can use a BLAS-style scatter-gather to
incorporate SIMD, which can improve cache performance and memory
throughput; this approach has the additional overhead for the gather
operation.  We find on a synthetic dataset in which we vary the
sparsity from 0.01 to 1.0, \Dense can be up to 2$\times$ faster than
\Sparse (for sparsity=1.0) while \Sparse can be up to 4$\times$ faster
than \Dense (for sparsity=0.01).

The dense vs. sparse tradeoff might change on newer CPUs with
\textsf{VGATHERDPD} intrinsic designed to specifically speed up the
gather operation. However, our current machines do not support this
intrinsics and how to optimize sparse and dense computation kernel is
orthogonal to the main goals of this paper.

\paragraph*{Row-major and Column-major Storage}
There are two well-studied strategies to store a data matrix $A$:
\Rowmajor and \Columnmajor storage.  Not surprisingly, we observed
that choosing an incorrect data storage \yell{strategy} can cause a
large slowdown. We conduct a simple experiment where we multiply a
matrix and a vector using row-access method, where the matrix is
stored in column- and row-major order. We find that the \Columnmajor
could resulting $9\times$ more L1 data load misses than using \Rowmajor
for two reasons: (1) our architectures fetch four doubles in a
cacheline, only one of which is useful for the current operation. The
prefetcher in Intel machines does not prefetch across page boundaries,
and so it is unable to pick up significant portions of the strided
access; (2) On the first access, the Data cache unit (DCU) prefetcher
also gets the next cacheline compounding the problem, and so it runs
$8\times$
slower.\footnote{\url{www.intel.com/content/dam/www/public/us/en/documents/manuals/64-ia-32-architectures-optimization-manual.pdf}}
Therefore, \dimm always stores the dataset in a way that is consistent
with the access method---no matter how the input data is stored

\begin{figure*}[t!]
\scriptsize
\centering
\begin{tabular}{c|ccc|ccc|cc}
\Xhline{4\arrayrulewidth}
                      &   \multicolumn{3}{c}{\bf Target Architecture}    & \multicolumn{3}{|c|}{\bf Target Application}  & \multicolumn{2}{c}{\bf Target Efficiency}        \\ 
                      &   Multicore   & NUMA (SMP)  & Distributed  &  Data Mining & Graph Mining & Gradient-based   &  Hardware   &  Statistical  \\
\hline
Jin et al.~\cite{Jin:2005:TKDE}                          &    &  \checkmark  &    &   \checkmark & \checkmark   &     & \checkmark   &    \\
Ghoting et al.~\cite{Ghoting:2007:VLDBJ}                 &    &  \checkmark  &    &   \checkmark &    &     & \checkmark   &    \\
Tatikonda et al.~\cite{Tatikonda:2009:VLDB}              &    &  \checkmark  &    &   \checkmark &    &     & \checkmark   &    \\
Chu et al.~\cite{Chu:2006:NIPS}                          & \checkmark   &    &    &    &  \checkmark  &  \checkmark & \checkmark   &    \\
Zaki et al.~\cite{Zaki:1999:ICDE}                        &    &  \checkmark  &    &   \checkmark &    &     & \checkmark   &    \\
Buehrer et al.~\cite{Buehrer:2006:ICDM}                  &    &  \checkmark  &    &    &  \checkmark    &     & \checkmark   &    \\
Buehrer et al.~\cite{Buehrer:2007:PPOPP}                 &    &   &  \checkmark  &   \checkmark &    &     & \checkmark   &    \\
Zaki et al.~\cite{Parthasarathy:2001:KAIS,Zaki:1997:KDD} &    &  \checkmark  &    &   \checkmark &    &     & \checkmark   &    \\
\hline
Tsitsiklis et al.~\cite{Tsitsiklis:1986:AC}              &    &    &    &    &    &  \checkmark   &    &  \checkmark  \\ 
Niu et al.~\cite{DBLP:conf/nips/RechtRWN11}              & \checkmark   &    &    &    &    &  \checkmark   &    &  \checkmark  \\ 
Bradley et al.~\cite{DBLP:conf/icml/BradleyKBG11}        & \checkmark   &    &    &    &    &  \checkmark   &    &  \checkmark  \\
\hline
GraphChi~\cite{Kyrola:2012:OSDI}                                                 & \checkmark  &  \checkmark  &    & \checkmark   &  \checkmark  &  \checkmark   & \checkmark & \\
GraphLab~\cite{DBLP:conf/uai/LowGKBGH10,DBLP:journals/pvldb/LowGKBGH12}                                                 &    & \checkmark   & \checkmark   &  \checkmark  & \checkmark   &  \checkmark   & \checkmark & \\
MLlib~\cite{Sparks:2013:ICDM}                                                    &    &    & \checkmark   &  \checkmark  &  \checkmark  &  \checkmark   & \checkmark & \\
\hline
\dimm                                                    &    &  \checkmark  &    &    &    &  \checkmark   & \checkmark & \checkmark\\
\Xhline{4\arrayrulewidth}
\end{tabular}
\caption{A Taxonomy of Related Work}
\label{fig:taxonomy}
\end{figure*}

\section{Extended Related Work}

We extend the discussion of related work.  We summarize in
Figure~\ref{fig:taxonomy} a range of related data mining work. A key
difference is that \dimm considers both hardware efficiency and
statistical efficiency for statistical analytics solved by first-order
methods.

\paragraph*{Data Mining Algorithms}
Probably the most related work is by Jin et
al.~\cite{Jin:2005:TKDE}, who consider how to take advantage of
replication and different locking-based schemes with different caching
behavior and locking granularity to increase the performance (hardware
efficiency performance) for a range of data mining tasks including
K-means, frequent pattern mining, and neural networks.  Ghoting et
al.~\cite{Ghoting:2007:VLDBJ} optimize cache-behavior of frequent
pattern mining using novel cache-conscious techniques, including
spatial and temporal locality, prefetching, and tiling. Tatikonda et
al.~\cite{Tatikonda:2009:VLDB} considers improving the performance
of mining tree-structured data multicore systems by decreasing the
spatial and temporal locality, and the technique they use is by
careful study of different granularity and types of task and data
chunking. Chu et al.~\cite{Chu:2006:NIPS} apply the MapReduce to a
large range of statistical analytics tasks that fit into the
statistical query model, and implements it on a multicore system and
shows almost linear speed-up to the number of cores. Zaki et
al.~\cite{Zaki:1999:ICDE} study how to speed up classification
tasks using decision trees on SMP machines, and their technique takes
advantage data parallelism and task parallelism with
lockings. Buehrer and Parthasarathy et al.~\cite{Buehrer:2006:ICDM}
study how to build a distributed system for frequent pattern mining
with terabytes of data. Their focus is to minimize the I/O cost and
communication cost by optimizing the data placement and the number of
passes over the dataset. Buehrer et al.~\cite{Buehrer:2007:PPOPP}
study implementing efficient graph mining algorithms over CMP and SMP
machines with the focus on load balance, memory usage (i.e., size),
spatial locality, and the tradeoff of pre-computing and re-computing.
Zaki et al.~\cite{Parthasarathy:2001:KAIS,Zaki:1997:KDD} study on
how to implement parallel associated rule mining algorithms on shared
memory systems by optimizing reference memory locality and data
placement in the granularity of cachelines.  This work also considers
how to minimize the cost of coherent maintenance between multiple CPU
caches. All of these techniques are related and relevant to our work,
but none consider optimizing first-order methods and the affect of
these optimizations on their efficiency.

\paragraph*{High Performance Computation} 
The techniques that we considered in \dimm for efficient implementation
(Section~\ref{sec:impl_details}) are not new, and they are borrowed
from a wide range of literature in high performance computation,
database, and systems.
Locality is a classical technique: worker and data collocation
technique has been advocated since at least
90s~\cite{Anderson:1993:PLDI,Carr:1994:ASPLOS} and is a common
systems design principle~\cite{Silberschatz:1991:Book}.

The role of dense and sparse computation is well studied in the by the
HPC community. For example, efficient computation kernels for
matrix-vector and matrix-matrix
multiplication~\cite{Bell:2008:NVIDIA,Bell:2009:SC,DAzevedo:2005:ICCS,Williams:2007:SC}. In
this work, we only require dense-dense and dense-sparse matrix-vector
multiplies. There is recent work on mapping sparse-sparse multiplies
to GPUs and SIMD~\cite{Yang:2011:VLDB}, which is useful for other
data mining models beyond what we consider here.

The row- vs. column-storage has been intensively studied
by database community over traditional relational database~\cite{Ailamaki:2001:VLDB}
or Hadoop~\cite{He:2011:ICDE}.
\dimm implements these techniques to make sure our
study of hardware efficiency and statistical efficiency
reflects the status of modern hardware,
and we hope that future development on these topics can
be applied to \dimm.

\paragraph*{Domain Specific Languages}

Domain specific languages (DSLs) are intended to make it easy for a
user to write parallel programs by exposing domain-specific
patterns. Examples of such DSLs include
Galois~\cite{Nguyen:2014:ASPLOS,Nguyen:2013:SOSP} and
OptiML~\cite{Sujeeth:2011:ICML} for
Delite~\cite{Chafi:2011:PPOPP}. To be effective, DSLs require the
knowledge about the trade-off of the target domain to apply their
compilation optimization, and we hope the insights from \dimm
can be applied to these DSLs.

\paragraph*{Mathematical Optimization}
Many statistical analytics tasks are mathematical optimization
problems. Recently, the mathematical optimization community \yell{has
  been} looking at how to parallelize optimization
problems~\cite{DBLP:conf/nips/RechtRWN11,DBLP:conf/nips/ZinkevichWSL10,Ji:2014:ICML}. For
example, Niu et al.~\cite{DBLP:conf/nips/RechtRWN11} for SGD and
Shotgun~\cite{DBLP:conf/icml/BradleyKBG11} for SCD. A lock-free
asynchronous variant was recently established by Ji et
al.~\cite{Ji:2014:ICML}.

\section{Additional Experiments}


\subsection{More Detailed Tuning Information for \spark}

We report details of how we tune our \spark installation
for fair comparison. Figure~\ref{fig:sparktuning}
shows the list of parameters that we used to tune
\spark. For each combination of the parameter, we
run one experiment for measuring the throughput using
parallel sum, and use it for all other experiments
to maximize the performance. For each task, we try
all combinations of step size and batch size.

\begin{figure}
\scriptsize
\begin{tabular}{ccc}
\Xhline{4\arrayrulewidth}
{\bf Type} & {\bf Parameters}   &    {\bf Values} \\
\hline
Statistical           &  Step size    & 100, 10, 1, 0.1, 0.01, 0.001, 0.0001 \\
Efficiency          &  Batch size   & 100\%, 50\%, 10\%, 1\% \\
\hline
          &  Data Replication & 1, 2, 3                 \\
          &  Serialization   & True, False              \\
Hardware          &  Storage Level   & MEMORY\_ONLY             \\
Efficiency          &  Compression    & True, False               \\
          &  locality.wait  & 1, 100, 1000, 3000, 10000      \\
          &  SPARK\_MEM    & 48g, 24g, 1g               \\
          &  numactl      & localloc, interleave, NA   \\
\Xhline{4\arrayrulewidth}
\end{tabular}
\caption{The Set of Parameters We Tried for Tuning \spark}
\label{fig:sparktuning}
\end{figure}

\paragraph*{Statistical Efficiency: Step Size and Batch Size}

We observe that step size and batch size of gradient
together has significant impact on the time that
\spark needs to converge. As shown in Figure~\ref{fig:sparktuning},
for each experiment, we try 28 different combinations
of these settings (7 step sizes and 4 batch sizes).
We see that these parameters could contribute to more than 
100$\times$ in the time to converge to the same loss 
on the same dataset! Therefore, as shown in
Figure~\ref{fig:sparktuning}, we tried a large range
of these two parameters and pick the best one to report.

\paragraph*{Sources of Overhead in \spark}

\spark has overhead in scheduling the task and provide fault
tolerance, both of which are features that \dimm does not support.  To
make our comparison as fair as possible, we conduct the following
experiments to understand how scheduling and fault tolerance impact
our claims.

We implement our own version of batch-gradient descent algorithm in
\dimm by strictly following \mllib's algorithm in C++. On Forest, we
first observe that our own batch-gradient implementation uses similar
numbers of epochs (within 5\%) to converge to 1\% loss as \mllib given
the same step size and batch size. Second, for each epoch, our
batch-gradient implementation is 3-7$\times$ faster cross different
architectures--this implies that \mllib does have overhead compared
with \dimm's framework. However, our own batch-gradient implementation
is still 20-39$\times$ slower than \dimm cross different
architectures.  

We break down the execution time into the number of epochs that each
system needs to converge and the time that \mllib used for scheduling
and computation.  In particular, we use the Forest dataset as an
example. On this dataset, \dimm uses 1 epoch to converge to 1\% loss,
while both \mllib and our own C++ implementation use 63 and 64 epochs,
respectively. \mllib uses 2.7 seconds for these 64 epochs, and 0.9
seconds of these are used for scheduling, and other 1.8 seconds are
used to enumerate each example, and calculate the gradient.\footnote{
  We observe similar break down on other datasets except the smallest
  dataset, Reuters. On this dataset, the time used for scheduling is
  up to 25$\times$ of the computation time.}  The difference in the
number of epochs to converge implies that the difference between
\mllib and \dimm is not caused by low-level implementations, instead,
that \mllib only implements a subset of points in \dimm's tradeoff
space.  

\paragraph*{Hardware Efficiency}

We summarize the impact of parameters to the
throughput of \mllib. For each out of totally 540 
combinations of all seven parameters related to 
hardware efficiency, we run the parallel sum 
to measure the throughput. We find,
not surprisingly, that the parameter
SPARK\_MEM has significant impact on the 
throughput--On Music, when this parameter is
set to 48GB, \spark achieves 7$\times$ speed-up over
1GB. This is not surprising because this parameter
sets the amount of RAM that \spark can use.
We also find that, given the SPARK\_MEM parameter
to be 48GB, all other parameters only have less than 50\%
difference with each other.
Therefore, in our experiments we always use SPARK\_MEM
and set other parameters to be the setting that
achieves highest throughput in our experiment
on the corresponding dataset.

\subsection{Comparison with Delite}

\begin{figure}[t!]
\centering
\includegraphics[width=0.3\textwidth]{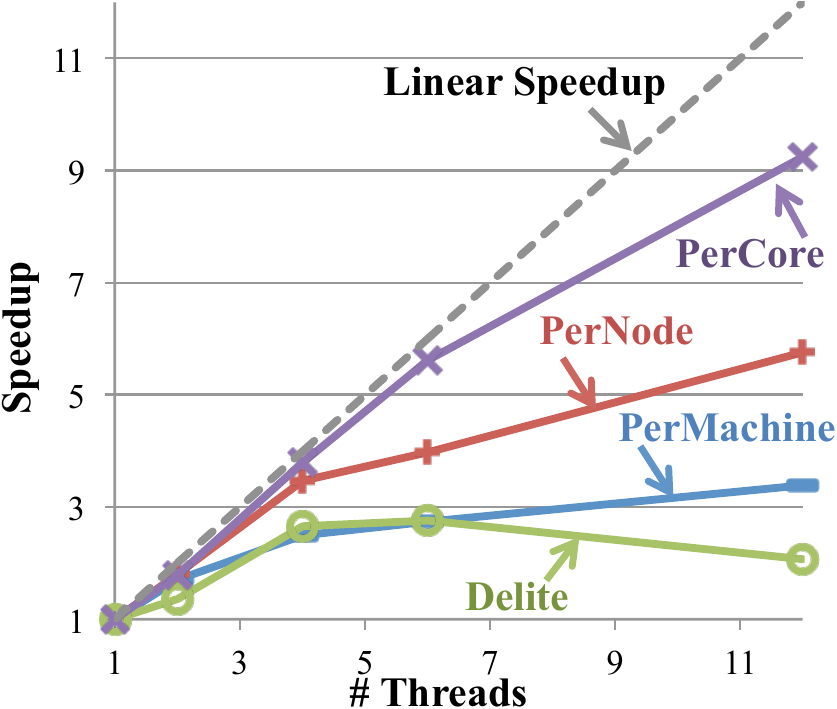}
\caption{Comparison with Delite using LR (Music) on local2.}
 \label{fig:delite}
 \end{figure}

Recently, there have been a trend of using domain
specific language to help user write parallel
programs more easily. We conduct a simple experiment
with one popular DSL, namely Delite~\cite{Chafi:2011:PPOPP},
to illustrate that the tradeoff we studied in this paper
has the potential to help these DSLs to achieve higher
performance and quality.

We use the official implementation of logistic regression
in Delite~\cite{Chafi:2011:PPOPP} and run both \dimm and
Delite on the Music dataset using local2. 
We try our best effort for the locality of Delite by
trying different settings for numactl.
We vary the
number of threads that each program can use and plot
the speed-up curve as shown in Figure~\ref{fig:delite}.

First, we see from Figure~\ref{fig:delite} that different
model replication strategy in \dimm has different
speed-up behavior. Not surprisingly, \percpu speeds up
more linearly than \pernode and \permachine. These observations
are consistent with the hardware efficiency that we discussed
in this paper. More interestingly, we see that Delite
does not speed-up beyond a single socket (i.e., 6 cores).
Therefore, by applying the \pernode strategy in DimmWitted
to Delite, we hope that we can improve the speed-up behavior
of Delite as we illustrated in Figure~\ref{fig:delite}.

\subsection{Scalability Experiments}

We validate the scalability of \dimm by testing it
on larger dataset.

\paragraph*{Dataset} We follow Kan 
et al.~\cite{Kan:2005:CIKM} to create a dataset
that contains 500 million examples, 100K features
for each example, and 4 billion non-zero elements by 
using a Web-scale data
set called ClueWeb.\footnote{http://lemurproject.org/clueweb09/} 
ClueWeb contains 500 million Web pages, and the 
approach of Kan et al. tries predict the PageRank
score of each Web page by using features from its URLs
by a least squares model.

\begin{figure}[t!]
\centering
\includegraphics[width=0.3\textwidth]{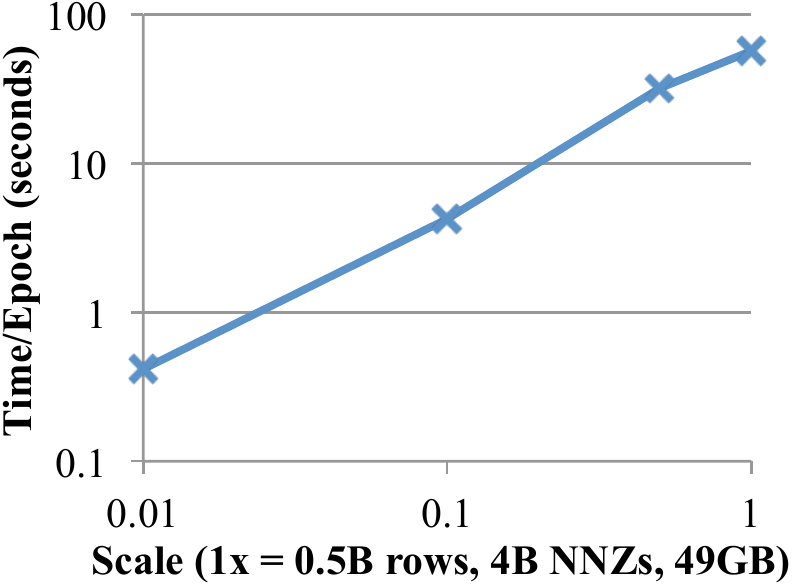}
\caption{Scalability of \dimm using ClueWeb 2009 on local2.}
 \label{fig:scalability}
 \end{figure}

\paragraph*{Result} To validate the scalability of 
\dimm, we randomly subsampled 1\% examples, 10\% examples,
and 50\% examples to create smaller datasets. We run
\dimm using the rule-of-thumbs 
in Figure~\ref{fig:plan_of_choice}, and measure the
time that \dimm used for each epoch.
Figure~\ref{fig:scalability} shows the result. We
see that on this dataset, the time that \dimm needs
to finish a single epoch grows almost linearly with the
number of examples. We believe that this is caused by
the fact that for all sub-sampled datasets and the
whole dataset, the model (100K weights) 
fits in the LLC cache.

\subsection{Importance Sampling as a Data Replication Strategy}

The \sharding and \fullreplication sampling
scheme that we discussed in Section~\ref{sec:optimizer}
assumes that data tuples are equally important.
However, in statistic analytics, it is not uncommon
that some data tuples are more important than others.
One example is the linear leverage score.

\begin{example}[Linear Leverage Score~\cite{drineas}]
For $A \in \R^{N \times d}$ and $b \in \R^{N}$.
Define $s(i) = a_i^{T}\left(A^TA\right)^{-1}a_i$,
where $a_i$ is the $i^{th}$ row of $A$.
Let $\tilde{A}$ and $\tilde{b}$ be
the result of sampling $m$ rows, where row $i$ is selected
with probability proportional to $s(i)$.
Then, {\em for all} $x \in \R^d$, we have
\[ \Pr\left[ \left|\|Ax - b\|_2^2 - \frac{N}{m} \|\tilde{A}x - \tilde{b}\|_2^2\right| < \varepsilon \|Ax - b\|_2^2 \right] > \frac{1}{2} \]
So long as $m > 2\varepsilon^{-2} d \log d$.

\end{example}

For general loss functions (e.g., logistic loss), the linear leverage
score calculated in the same way as above does not necessarily satisfy
the property of approximating the loss.  However, we can still use
this score as a {\em heuristic} to decide the relative importance of
data examples.  In \dimm, we consider the following protocol that we
called \textsf{Importance}.
Given a dataset $A$,
we calculate the leverage score $s(i)$ of the $i^{th}$
row as $a_i^T(A^{T}A)^{-1}a_i$.
The user specifies the error tolerance
$\epsilon$ that is acceptable to her, and
for each epoch, \dimm samples for each worker
$2\varepsilon^{-2} d \log d$ examples
with a probability that is propositional
to the leverage score.
This procedure is implemented in \dimm
as one data replication strategy.

\begin{figure}[t!]
\centering
\includegraphics[width=0.35\textwidth]{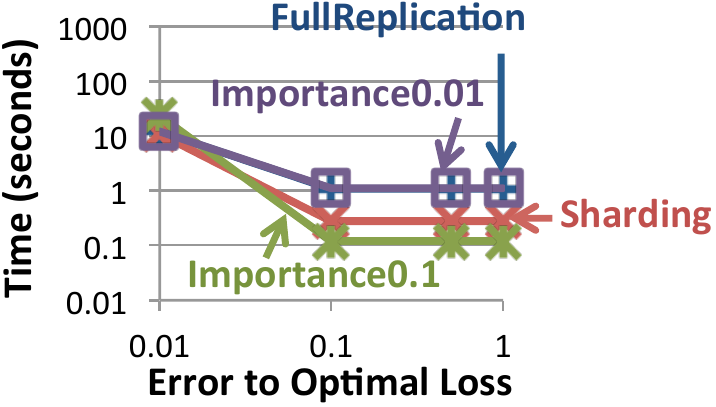}
\caption{Important Sampling on Music (local2).}
\label{fig:importance}
\end{figure}

\paragraph*{Experimental Results}

We run the above importance sampling on the same data
set as Section~\ref{sec:experiment}, and validate that
on some datasets the importance sampling scheme can
improve the time that \dimm needs to converge
to a given loss. Figure~\ref{fig:importance}
shows the results of comparing
different data replication strategies on Music running 
on local2, where Importance0.1 and Importance0.01
uses 0.1 and 0.01 as the error tolerance $\epsilon$, 
respectively.

We see that, on Music, Importance0.1 is 3x faster 
than FullReplication, for 10\% loss. This is caused 
by the fact that Importance0.1 processes only 10\% of 
the data compared with FullReplication. However, 
Importance0.01 is slower than FullReplication. This 
is because when the error tolerance is lower, the number
of samples one needs to draw for each epoch increases.
For Music, Importance0.01 processes the same
amount of tuples than FullReplication.

\section{Detailed Description of Extensions}

We describe in more details of each extension that
we mentioned in Section~\ref{sec:extension}.

\begin{figure}[t!]
\centering
\includegraphics[width=0.5\textwidth]{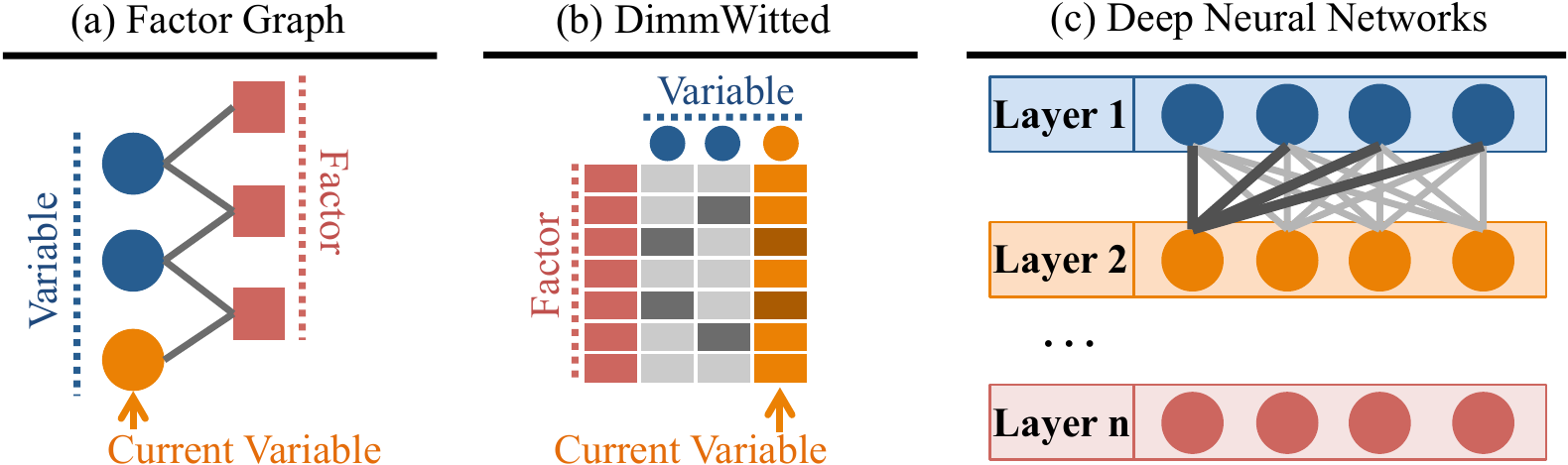}
\caption{Illustration of Factor Graph and Deep Neural Networks
in \dimm. (a) and (b) show a factor graph and how \dimm represents
it as column-to-row access. (c) shows a deep neural network,
and the de facto approach to solve it is to run SGD for
each layer \dimm in a round-robin fashion.}
\label{fig:factorgraph}
\end{figure}

\subsection{Gibbs Sampling}

Figure~\ref{fig:factorgraph}(a) illustrates
a factor graph, which is a bipartite graph that
contains a set of variable, a set of factors, and a set of
links between variables and factors.
To run Gibbs sampling over a factor graph,
one processes one variable at a time to calculate
the conditional probability for different
assignment of this variable. This involves fetching
all connected factors and all current assignments
of variables that connected to these factors.
Gibbs sampling then update the current variable
assignment by randomly sampling a value according
to the conditional probability and proceed to
the next random variable.
Similar to first order methods, recent theory
proves a lock-free protocol to sample multiple
variables at the same time~\cite{Johnson:2013:NIPS}.
We also know from classic statistical theory~\cite{Robert:2005:Book}
that one can maintain multiple copy of the same
factor graph, and aggregate the samples produced
on each factor graph at the end of execution.

Figure~\ref{fig:factorgraph}(b) illustrates
how \dimm models Gibbs sampling as column-to-row access.
We see that each row corresponding to one factor,
each column corresponding to one variable,
and the non-zero elements in the matrix correspond
to the link in the factor graph.
To process one variable, \dimm fetches one column
of the matrix to get the set of factors, and
other columns to get the set of variables that
connect to the same factor.

In \dimm, we implement the \pernode strategy
for Gibbs sampling by running one independent 
chain for each NUMA node. At the end of sampling,
we can use all samples generated from each NUMA node
for estimation. Therefore, we use throughput, i.e.,
number of samples generated per second as the measurement
for performance in 
Section~\ref{sec:extension}.\footnote{There has been a long
historical discussion about the tradeoff between a single deep
chain and multiple independent chains in statistics. This
tradeoff is out of the scope of this paper.}
In \dimm, we implement Gibbs sampling for general factor
graphs, and compare it with one hand-coded implementation
for topic modeling in \graphlab. We run all
systems on local2 with 100K documents and 20 topics.
We find that on local2,
\dimm's implementation is 3.7$\times$ faster than
\graphlab's implementation without any application-specific
optimization.



\subsection{Deep Neural Networks}

Figure~\ref{fig:factorgraph}(c) illustrates a Deep Neural Network
as we described in Section~\ref{sec:extension}.
Stochastic gradient descent is the de facto algorithm
to solve a neural network~\cite{LeCun:1998:PIEEE}, with
one twist that we will discuss as follows.
As shown in Figure~\ref{fig:factorgraph}(c), a
deep neural network usually contains multiple layers,
and the SGD algorithm needs to be run within each layer,
and process all layers in a round-robin fashion.
Therefore, in \dimm, we use the same SGD code path inside
each layer one at a time, and invoke this code path multiple times
to process different layers.
\balance


\end{document}